\documentclass[structabstract]{aa}  

\usepackage{txfonts}
\usepackage{graphicx}

\usepackage{natbib}
\bibpunct{(}{)}{;}{a}{}{,}   % to follow the A&A style

\usepackage{longtable,lscape}

\newcommand{\mjup}{M$_{\mbox{\tiny Jup}}$}

\newcommand{\ms}{m\,s$^{-1}$}

\newcommand{\msun}{M$_{\odot}$}

\newcommand{\iotadra}{$\iota$~Dra}

\newcommand{\fpl}{$f_{\mbox{\tiny planet}}$}

\newcommand{\fpc}{$f_{\mbox{\tiny planet+cand}}$}

\newcommand{\np}{$n_{\mbox{\tiny planet}}$}
\newcommand{\nc}{$n_{\mbox{\tiny cand}}$}
\newcommand{\ns}{$n_{\mbox{\tiny stars}}$}

\newcommand{\bfa}{}

\newcommand\phn{\phantom{0}}% 

\begin{document}

   \title{Precise Radial Velocities of Giant Stars}

   \subtitle{VII. Occurrence Rate of Giant Extrasolar Planets as a Function of Mass and 
             Metallicity\thanks{Based on observations collected at Lick Observatory,
             University of California}}

   \author{Sabine Reffert\inst{1}
          \and
          Christoph Bergmann\inst{1,2}
          \and
          Andreas Quirrenbach\inst{1}
          \and
          Trifon Trifonov\inst{1,3}
          \and 
          Andreas K\"unstler\inst{1,4}
          }

   \institute{Landessternwarte, Zentrum f\"ur Astronomie der Universit\"at Heidelberg, K\"onigstuhl 12,
              69117 Heidelberg, Germany, 
              \email{sreffert@lsw.uni-heidelberg.de}
         \and
             University of Canterbury, Department of Physics and Astronomy, Christchurch 8041, New Zealand
         \and
             Department of Earth Sciences, The University of Hong Kong,
             Pokfulam Road, Hong Kong
         \and
             Leibniz-Institut f\"ur Astrophysik Potsdam, An der Sternwarte 16, 14482 Potsdam, Germany   
             }

   \date{Received xxxxx xx, xxxx; accepted xxxxx xx, xxxxx}

% \abstract{}{}{}{}{} 
% 5 {} token are mandatory
 
  \abstract
  % context (optional)
   {We have obtained precise radial velocities for a sample of 373 G and K type giants
    at Lick Observatory regularly over more than 12 years. Planets have been identified
    around 15 of these giant stars, and an additional 20 giant stars host planet candidates.}
  % aims (mandatory)
   {{\bfa We are interested in the occurrence rate of substellar companions around giant stars
    as a function of stellar mass and metallicity. We probe the stellar mass
     range from approximately 1 to beyond 3\,\msun, which is not being explored
     by main-sequence samples.}}
  % methods (mandatory)
    {{\bfa We fit the giant planet occurrence rate as a function
    of stellar mass and metallicity with a Gaussian and an exponential
    distribution, respectively.}}
  % results (mandatory)
    {We find strong evidence for a planet-metallicity correlation among the secure
     planet hosts of our giant star sample,
     in agreement with the one for main-sequence stars. However,
    the planet-metallicity correlation is absent for our sample of
    planet candidates, raising the suspicion that a good fraction of them might indeed
    not be planets despite clear periodicities in the radial velocities. 
    {\bfa Consistent with the results obtained by Johnson
    for subgiants, the giant planet occurrence rate increases in the stellar mass
    interval from 1 to 1.9\,\msun. However, there is a maximum at
    a stellar mass of $1.9^{+0.1}_{-0.5}$\,\msun, and the occurrence rate} drops 
    rapidly for masses larger than 2.5--3.0\,\msun. 
    We do not find any planets around stars more massive than
    2.7\,\msun, although there are 113 stars with masses between 2.7 and
    5\,\msun\ in our sample (corresponding to a giant planet occurrence rate
    smaller than 1.6\% at 68.3\% confidence in that stellar mass bin). We also show
    that this result is not a selection effect related to the planet detectability
    being a function of the stellar mass.}
    {We conclude that giant planet formation or inward migration
     is suppressed around higher mass stars,
     possibly because of faster disk depletion coupled with a longer migration timescale.}


   \maketitle
%
%________________________________________________________________

\section{Introduction}
About {\bfa 1500} confirmed extrasolar planets are known today
{\bfa (\citealt{wright11}\footnote{\texttt{http://www.exoplanets.org/}})}. 
{\bfa The rate of detection and confirmation of exoplanets has increased
dramatically in the recent past, with almost one third of exoplanets confirmed
in early 2014 alone \citep{rowe14,lissauer14} and another third being discovered
only during the last two years.}
Roughly two thirds
of all confirmed planets have been found with the transiting method,
{\bfa and many more candidates are known, especially those discovered with the}
{\bfa {\it Kepler} space telescope
(see \citealt{borucki11a,borucki11b,batalha13} for Kepler planet candidates and 
\citealt{howard12,fressin13} for statistical analyses of planet occurrence rates)}.
The radial velocity (RV) technique, which {\bfa accounts for the detection of the
remaining third of confirmed planets}, has major difficulties
with the detection of planets around stars more massive than 
$\approx 1.5\,\mathrm{M_{\sun}}$, since hotter stars have fewer 
absorption lines and also rotate more rapidly, which both adversely affect 
the radial velocity precision
{\bfa \citep{galland05a,lagrange09}. A few giant planets and brown dwarfs have 
been found around early dwarf 
stars despite these difficulties \citep[see e.g.{}][]{galland05b,galland06}}.

In order to avoid these difficulties one can target moderately massive stars 
that have already left the main sequence and have evolved into giant stars. 
As giant stars are
cooler than their predecessors on the main sequence and have slower rotation
rates, their spectral lines are more numerous and also less
broadened, so that one can measure their radial velocities very precisely. 

While the statistics on properties of extrasolar planets around main-sequence
stars has continuously improved throughout the last decade, the number of
known substellar companions around giant stars is still comparatively small. 
So far, {\bfa 63} companions in {\bfa 59} systems have been 
announced\footnote{We maintain a list of planet discoveries around giant stars at 
http://www.lsw.uni-heidelberg.de/users/sreffert/giantplanets.html}.
They have emerged from a variety of Doppler surveys, including our own using
the Hamilton Spectrograph at Lick Observatory 
\citep{frink01,frink02,reffert06,mitchell13,trifonov14} as
well as surveys conducted with FEROS at La Silla Observatory \citep{setiawan03,setiawan05}, 
with the HIDES spectrograph at Okayama Observatory
\citep{sato03,sato07,sato08a,sato08b,sato10,sato12}, Tautenburg 
\citep{hatzes05,doellinger07,doellinger09a,doellinger09b}, the Penn State
Tor\'un planet search with the Hobby-Eberly Telescope 
\citep{niedzielski07,niedzielski09a,niedzielski09b,gettel12a,gettel12b},
the BOES spectrograph at Bohyunsan Observatory
\citep{han10,lee12a,lee12b,lee13}{\bfa , and the search 
by \citet{jones11,jones13,jones14,jones14b}, which employs FEROS at La Silla
Observatory as well as the FECH and CHIRON spectrographs at 
Cerro Tololo Inter-American Observatory}. 

The planet-metallicity correlation \citep{fischer05,udry07} is by now well established 
for main-sequence stars, indicating that a star with high metallicity has a much
higher probability of hosting a planet than a star with a lower metallicity. 
For subgiants, the same trend with metallicity was observed by \citet{johnson10}, 
although it is not as strong as for main-sequence stars. 
In addition to metallicity, \citet{johnson10} also found
a correlation between planet occurrence rate and stellar mass in the sense
that higher mass stars have a higher probability of hosting a giant planet.
In the case of planet-hosting evolved stars it is
not clear yet whether a correlation between planet occurrence rate and
metallicity exists. While \citet{pasquini07}, \citet{takeda08} {\bfa and \citet{mortier13}}
do not find
evidence for such a correlation in their samples of G and K giants, 
\citet{hekker07} find indications for a positive correlation between metallicity
and planet occurrence rate for K~giant stars. However, one has to keep in mind
that the number of securely established planets around evolved stars originating from 
a single homogeneously selected and observed sample was still quite small at
that time.

Recently, \citet{maldonado13} have reinvestigated a possible link between
metallicity and giant planet occurrence rate in a sample of published subgiant 
and giant planet hosts, using a control sample of nearby Hipparcos giants
(which were not actually searched for planets and which is certainly not ideal - 
given the huge number of bright giant stars slightly different selection criteria
can result in a sample with different characteristics). \citet{maldonado13}
found no planet-metallicity correlation for giant stars with stellar masses smaller
than 1.5\,\msun, but did find a positive planet-metallicity correlation
for giant stars with masses larger than 1.5\,\msun\ and for subgiants.
The issue of a planet-metallicity correlation for giant stars is thus far
from settled. 

In this paper we search for correlations between planet occurrence
rate and either stellar metallicity and/or stellar mass in our sample of 
373 G and K giants,
which we have been monitoring at Lick Observatory since 1999.
In particular, stellar masses in our parent sample range from about 1 to 5\,\msun\
and thus extend significantly beyond the mass range of the combined main-sequence and 
subgiant sample from \citet{johnson10}, which covers the mass range from 0.2 to 2\,\msun.
{\bfa Our detection capability of planetary companions is similar to that used in
other studies: for about 60\% of our monitored stars, we are sensitive to 
planets with an RV semi-amplitude of 30~m/s and a period less than 4~years, which
corresponds to the detection threshold used by \citet{fischer05}.} 

According to the usual definition involving deuterium burning,
one would have to call objects with masses of more than about 13\,\mjup\
\citep{spiegel11} brown dwarfs rather than planets. 
However, taking also the formation scenario into account,
one might want to include objects in the mass range from roughly 13 to 
25--30~\mjup\ among the planet category, especially those which formed 
around stars more massive than 1~\msun\ 
{\bfa \citep[see e.g.{}][]{baraffe10,schneider11}.}
We will not make a distinction
at the deuterium burning mass, but rather call all these objects planets
(or planet candidates) for simplicity.

The paper is organized as follows. 
In Section~2 we describe the Doppler survey at Lick Observatory. 
Section~3 is dedicated to the characterization of the stellar sample and 
in particular describes
how we derive stellar masses for our sample of giant stars. The planets that
we have found in our sample are discussed in Section~4. In Section~5 we statistically
analyze planet occurrence rate as function of stellar metallicity and
stellar mass and present our results. Section~6 provides a detailed 
discussion of our results including a comparison with planet formation models,
and finally we give a short summary in Section~7.

\section{Observations}
Since 1999, we have been carrying out a radial velocity survey of G and K giants
at UCO/Lick Observatory using the 0.6\,m Coud\'e Auxiliary Telescope
(CAT) together with the Hamilton Echelle Spectrograph {\bfa with a theoretical resolution
of approximately 60\,000 \citep{vogt87}; in practice we measure a resolution of
about 50\,000 at a wavelength of 6000~\AA.}
See \cite{frink01,frink02,reffert06} {\bfa for a description of our survey and earlier results.}
We follow the method for acquiring and reducing
data {\bfa using the iodine cell method} as described by \citet{butler96} 
and achieve a typical RV precision of about
5--8\,\ms\ with integration times of less than 30 minutes. 
We typically have accumulated between 20 and 100
observations per star, depending on brightness and observing
priority. 

Giant stars exhibit solar-like oscillations \citep[see e.g.{}][]{zechmeister08},
which manifest themselves as RV jitter in our observations due to the low
observing cadence. The amount of RV jitter depends on spectral type
\citep{frink01,hekker06}, with typical values of below 10--20~\ms\ for 
early K~giants and 
up to 50\,\ms\ or more for late K~giants and early M~giants. 
Thus, an RV precision of 5--8\,\ms\ is adequate for our survey.

\section{Stellar Sample}

\subsection{Sample Selection Criteria}
The selection criteria for the stars have been described by
\citet{frink01}. The stars are all brighter than 6$^{\mbox{\tiny th}}$\,mag in 
$V$ and have declinations between $-30^{\circ}$ and $+68^{\circ}$. The original 
sample only consisted of 86 K giants that were supposedly neither variable nor part of
a multiple system{\bfa; we especially checked a number of flags in the Hipparcos
Catalogue in order to make sure that there are no indications for so far
unresolved companions \citep[see appendix in][]{frink01}.}
Another 96 stars were included one year later with {\bfa basically the same
selection criteria; the only difference was that for the newly added stars}
photometric stability was not required anymore {\bfa (the coarse variability
flag in the Hipparcos Catalogue was allowed to be non-zero). However, it turned
out that the actual level of photometric variability is not very different between
the two parts of the sample; photometric variability is usually smaller than
0.01~mag and basically insignificant, so that we treat these two parts of the sample
together.}
Three of the stars were excluded from the sample when they were found
to be visual binaries. 

In 2004 the survey was extended again by 194 G and K
giants that have, on average, higher masses and bluer color $(0.8 \leq B-V
\leq 1.2)$. Bluer color means less intrinsic RV jitter \citep{frink01,hekker06},
which was the reason for the {\bfa inclusion} of late G~giants. 
{\bfa The selection of stars was performed in the following way. In a first
step, we identified all the G and K giant stars in the Hipparcos Catalogue
with relevant position, apparent magnitude and color (see above) via 
their reduced proper motion diagram, which nicely separates giant and dwarf 
stars based on their distinct kinematics. For this sample, we estimated
masses from evolutionary tracks \citep{girardi00} assuming solar metallicity.
In the end, we chose those stars with the highest masses derived in this way.
Binaries were also avoided as much as possible, but some have slipped in 
nevertheless.}

The reason for preferentially selecting more massive stars was
to test whether or not more massive stars also host more massive companions. In
retrospect we realized that these selection criteria for this sample were not
ideal for the issues addressed in this paper. As we did not have metallicities
on hand for these stars, the stellar masses were initially derived assuming
solar metallicity. However, the derivation of the masses used in this paper
does take metallicity into account, as described below. 
 
\subsection{Stellar Masses}

The stellar masses have been determined using a trilinear interpolation between
evolutionary tracks and isochrones \citep{girardi00}, respectively, 
and metallicity as a third parameter \citep{kuenstler08}. Because the location of
evolutionary tracks and isochrones depends on the assumed metallicity,
taking into account the metallicities is crucial. Most of the metallicities were derived
from the equivalent widths of iron lines by \citet{hekker07}. We performed
the interpolation for separate stages of the evolutionary tracks and isochrones
in order to determine the probabilities for the stars to be on the red giant
branch (RGB) or on the horizontal branch (HB), respectively. For each star we
generated 10\,000 positions within the three-dimensional space 
$(B-V,\,M_{\mathrm{V}},\,\left[\mathrm{Fe/H}\right])$, using the measured color,
absolute magnitude and metallicity and taking Gaussian errors into account. For
each of these generated positions we determined the stellar parameters from
evolutionary tracks by interpolation. The results have been weighted by the
initial mass function {\bfa (IMF)},
\begin{equation}
\mathrm{d}N \propto M^{-\alpha} \mathrm{d}M \quad,
\label{eq:imf}
\end{equation}
where we adopted a value of $\alpha = 2.35$ from \citet{imf}.
In addition {\bfa to the IMF}, the evolutionary timescale at a given position in the 
color-magnitude diagram (CMD),
\begin{equation}
v = \frac{\sqrt{\left(\Delta\left(B - V \right)\right)^2 + \left(\Delta M_{\mathrm{V}}\right)^2}}{\Delta t}
\end{equation}
has been taken into account {\bfa for the weighting. $v$ is the velocity in the CMD, at the 
position indicated by the star's $B-V$, $M_V$ and [Fe/H] values, or at the generated Monte
Carlo position in the CMD which is consistent with the observational errors in those quantities.}
This means that the probability for a star to be
located at a certain position in the CMD is lower the faster 
its evolution progresses at that particular position in the CMD. {\bfa Taken together, the weighting
with the IMF and the evolutionary timescale during the red giant phase closely resembles 
weighting with a present-day mass function for giant stars.}
Counting the successful interpolations and weighting
the results in the above-mentioned way yields the probabilities {\bfa and masses} of a star being
either on the RGB or on the HB. The masses {\bfa assuming either RGB or HB evolutionary status} are 
{\bfa mostly very similar to each other, with significant differences only in a few cases.} 
We provide {\bfa our derived} masses with errors and probabilities
for both {\bfa evolutionary} cases in Table~\ref{allstars}. In this paper we simply use the
mass which has the higher probability.

\begin{figure}
\resizebox{\hsize}{!}{\includegraphics{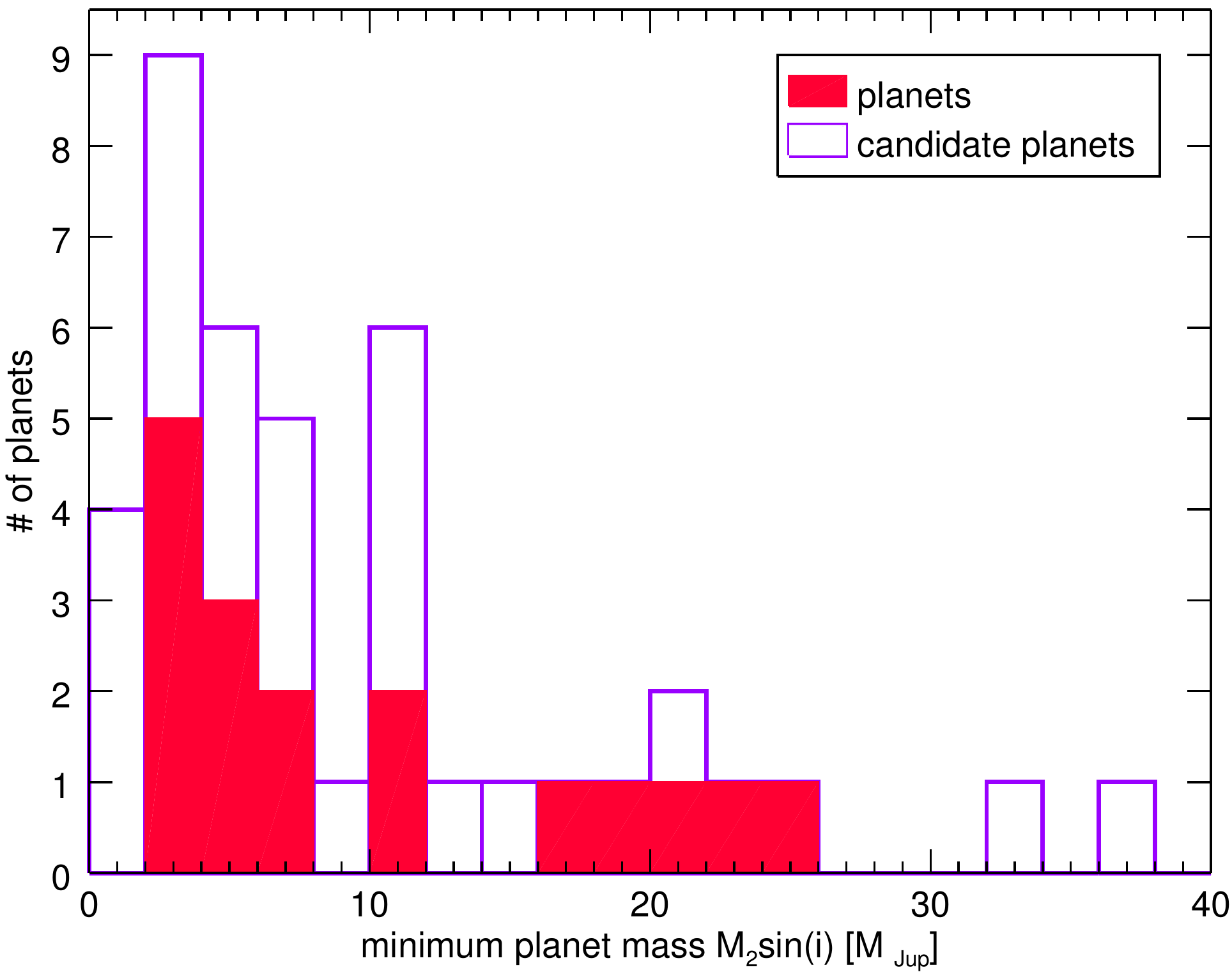}}
\caption{Histogram of minimum masses derived for 17 secure and
26 planet candidates in our Lick sample. Two of the 26
planet candidates have minimum masses larger than 40\,\mjup\ and are not 
shown in the plot.}
\label{masshist}
\end{figure}

\begin{figure*}
\sidecaption
\includegraphics[width=12cm]{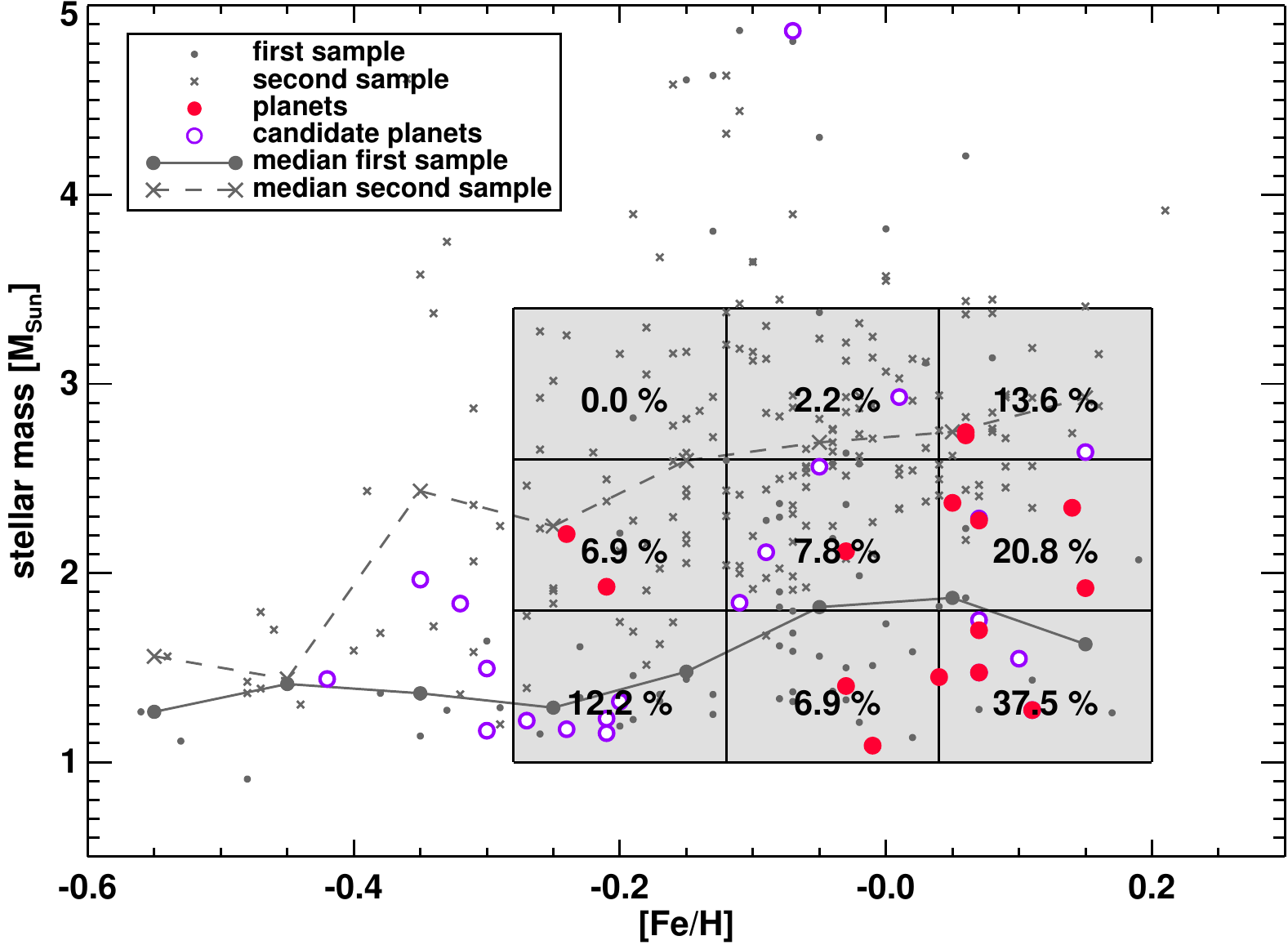}
\caption{Planet occurrence rate as a function of metallicity and stellar
mass in our Lick sample. The percentage numbers give the fraction
of the number of stars with a planet and/or planet candidate of all 
stars in that bin. One can clearly see that planet occurrence rate in our
Lick sample increases strongly with stellar metallicity
and decreases with stellar mass.
}
\label{fehmass}
\end{figure*}

Recently, \citet{lloyd11,lloyd13} and \citet{schlaufman13} have argued that masses derived
from evolutionary models for subgiant and giant stars are systematically too large. 
The argument from \citet{lloyd11,lloyd13} rests on the comparison of the
{\bfa mass distribution of planet host subgiants with a mass distribution 
constructed by integrating stellar model isochrones, 
taking the IMF, metallicity distribution and star formation history into account.}
\citet{schlaufman13} argue that the distribution of galactic space motions of subgiant
and giant planet hosts shows that the population has been dynamically heated and is
thus of lower mass than assumed so far. 
\citet{johnson13} point out that the discrepancy in {\bfa the mass distributions 
of subgiant planet hosts and model mass distributions based on galactic population 
synthesis} may be
{\bfa the} result of selection effects in the planet host sample, and conclude that the
masses derived on the basis of evolutionary tracks might still be correct.

{\bfa Similarly, there are many selection effects biasing the masses in our K giant
sample, and one should not expect that the masses follow those in an unbiased 
galactic model. Specifically, our Lick sample consists of three subsamples with 
different selection criteria. The sample is neither magnitude- nor volume-limited,
but has been selected to contain the most photometrically and astrometrically stable
stars available, based on a number of flags in the Hipparcos Catalogue. Specifically,
one of the subsamples was selected to contain the highest mass G and K giants,
in order to better test planet occurrence rate as a function of mass. Thus,
the statistical tests of \citet{lloyd11,lloyd13} and \citet{schlaufman13} do
not necessarily apply to our sample.
}

We simply note that while it is possible that our masses suffer from systematic
errors in the underlying evolutionary tracks, the masses for our Lick sample
would all be affected in the same way, {\bfa provided that the systematic error
shifts a particular evolutionary track in the CMD, but does not significantly 
affect its shape. This implies} that a comparison of the properties
of the higher mass stars of the sample to those of the lower mass stars in the
sample is still possible and not affected by an incorrect mass scale. 

{\bfa An independent test would be to determine asteroseismic masses for some of
the evolved stars in order to test spectroscopically determined masses; for some stars
in our Lick sample, the Kepler K2 mission will provide the necessary data. 
\citet{johnson14} compare mass determinations for the subgiant HD~185351, 
and find general agreement between the asteroseismic and
spectroscopic mass at ~1.8\,\msun; only once asteroseismology is combined with 
interferometry, the resulting mass estimate would be smaller by 2.6\,$\sigma$. 
Asteroseismic masses are available for two stars in our Lick sample: the
planet-bearing stars $\iota$~Dra \citep{baines11} and $\beta$~Gem \citep{hatzes12}.
In both cases, the asteroseismic masses are compatible with the spectroscopic masses
at a level of about 1-1.5\,$\sigma$; for $\iota$~Dra the asteroseismic mass
is the larger one, for $\beta$~Gem the spectroscopic one. A much larger sample
would be needed to unveil any systematic differences between the various methods
to derive stellar masses.}

\subsection{Stellar Parameter Table}
Besides stellar mass, the interpolation also yields other stellar parameters,
namely integrated mass loss $M_{\mathrm{loss}} $ (as compared to the zero age 
main-sequence masses), 
age $t_*$, effective temperature $T_{\mathrm{eff}}$, luminosity $L$,
radius $R_*$ (computed from the two previous quantities) and surface gravity
$\log g$. Table~\ref{allstars} provides these stellar parameters 
along with their formal errors as well as the $\left[\mathrm{Fe/H}\right]$ 
values for each star.

\section{Substellar Companion Statistics}
\label{sample}

Our goal is to investigate giant planet occurrence as a function of stellar mass
and metallicity. In order to do so, we need to identify all 
potential planets that have emerged from our Lick RV survey of G and K giant stars.

Since giant stars {\bfa show larger RV jitter} than main-sequence stars, it is
sometimes difficult to distinguish between an RV signal caused by a companion and
RV variations with an intrinsic stellar origin. 
In particular, non-radial pulsations could have similar periods as the planets 
we typically observe in our sample. {\bfa Non-radial pulsations have been identified
in giant stars on the basis of both {\it Corot} and {\it Kepler} data 
\citep[see e.g.{}][]{deridder09,carrier10,bedding10,stello13}, but it is not 
clear at present whether such single-mode, long-lived non-radial pulsations 
with high radial order \citep{hatzes99} as would be required to match the 
observed radial velocities really exist in giant stars.}
 
Radial pulsations are not a concern since they have much shorter 
timescales{\bfa, confirmed recently for red giants in large numbers via
asteroseismology performed with Kepler data, such as in \cite{bedding10},
\cite{kallinger12}, or \cite{stello13}}; they manifest themselves as stellar jitter 
in our RV curves.

Likewise, stellar spots which could in principle give rise to radial velocity
variations with the stellar rotation period can be excluded as the major contribution
to the observed radial velocity variability, since the spots would
have to be too large to be consistent with the low level of photometric variability
observed by Hipparcos. {\bfa This was shown explicitly to be the case for the planets
orbiting the K giants $\tau$~Gem and 91~Aqr in \cite{mitchell13} and for the interacting
planets around the K~giant $\eta$~Cet \citep{trifonov14}, and also applies
to all other planets and planet candidates in the Lick sample; none of the 
observed radial velocity variations is consistent with being caused by a spotted
star, in particular due to the very low level of photometric variability. 

We verified this again for all the planets and planet candidates discussed here, 
applying two tests. For the first test, we derived the spot filling
factor which would be required to generate the observed RV amplitude using the
relation from \cite{hatzes02}, estimated the resulting photometric variability
by factoring in the geometry and compared this to the photometric 
variability observed by Hipparcos. In many cases, in particular for those companions with
short periods and consequently large RV amplitudes, this resulted in spot filling factors 
larger than 5--10\%, 
which cannot be brought in line with the small photometric variations observed by Hipparcos.
For the second test, we assumed that the RV periods which we observe match the rotation period, and
calculated the resulting rotational velocity $v_{\mathrm{rot}}$ utilizing the derived stellar radii from
Table~\ref{allstars}. We compared this to the measured projected rotational velocities
$v\cdot\sin i$ from \cite{hekker07}. In most cases the rotational velocities $v_{\mathrm{rot}}$ 
are significantly smaller than the projected rotational velocities, which is not possible. 
This applies specifically to the stars with longer periods and thus smaller
rotational velocities $v_{\mathrm{rot}}$, increasing the discrepancy. 
In other words, the periods which we see in the RV data are typically larger than the 
rotational periods of the stars. These tests clearly showed that rotational modulation
of stellar features is not a viable explanation for the observed RV variability.}

The typical level of intrinsic radial velocity variation scales with $B-V$ 
\citep{frink01} and/or surface gravity \citep{hekker08};
while the expected stellar jitter of our late G and early K giants is less than
approximately 20~\ms, the jitter for late K giants (with $B-V$ around 1.6)
is up to 100~\ms\ and more. As a result, smaller mass or longer period planets,
which have smaller RV semi-amplitudes, cannot be identified as easily as those
around main-sequence stars. One needs more observations to compensate for
the stellar jitter, and longer time baselines covering several periods 
to ensure that phase and amplitude of the observed RV signal are constant 
over time. 

\begin{figure}
\resizebox{\hsize}{!}{\includegraphics{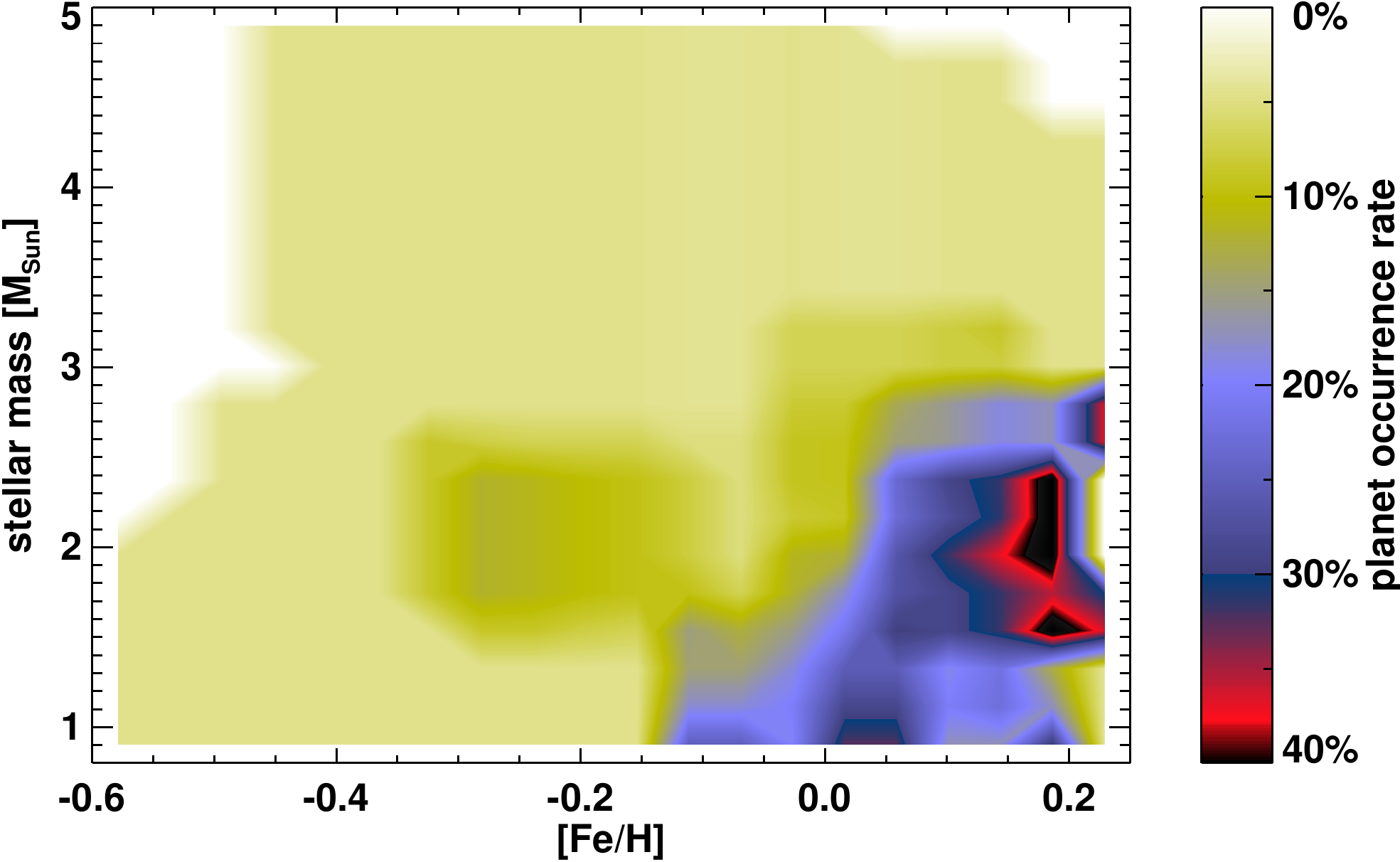}}
\caption{Planet occurrence rate (color coded) as a function of metallicity 
and stellar mass in our Lick sample; only the secure planets have been 
considered. Data have been heavily smoothed, so that the general trend
in planet occurrence rate becomes apparent.
There is a clear maximum in planet occurrence rate for metallicities
of about 0.2 and masses of about 2\,\msun.
}
\label{fehmassmap}
\end{figure}

For these reasons we have divided the tentative detections present in our
Lick sample into two categories: planets and planet candidates. Planets are
those which we consider secure discoveries, and planet candidates are those
whose existence is more doubtful. Planets usually have a clear and convincing
periodic RV signal, and have been observed over several cycles. 
Planet candidates also have clear periodic signals (highly significant
in Lomb-Scargle periodograms), but the ratio between RV semi-amplitude and 
stellar jitter is smaller than for the secure planets, so that stellar 
jitter becomes more prominent in the RV curve. 

Distinguishing
planet candidates from non-radial pulsations is rather difficult and 
requires many observations over a long time baseline in order to test
for phase and amplitude stability of the RV signal. Another possibility
to distinguish between pulsations and an orbiting companion is
the comparison of the RV signal computed from different wavelength regions
of the spectrum. We have started observing 20 K~giants with periodic
radial velocity patterns from our Lick sample with the CRIRES infrared 
high resolution spectrograph at VLT. These observations should ultimately 
tell us whether the periodic RV pattern is due to a companion (same pattern in 
the Lick and CRIRES velocities) or not (different amplitude and/or phase in
Lick and CRIRES velocities), after we have accumulated observations
over several years. 

In our Lick sample, we have identified secure planets around 15 stars.
Two of these have an additional secure planet and two an additional
planet candidate. We found planet candidates around 20 stars; four of
these have an additional planet candidate. Two stars display a
linear trend in the radial velocities on top of the planet candidate periodicity,
which could be indicative of substellar objects in very long orbits,
but are not considered here further. Two secure planets and four
planet candidates are found in spectroscopic binaries.
Altogether, this adds up to 43 (candidate) planets in 35 systems. 

A histogram of minimum planet masses is shown in Fig.~\ref{masshist}. 
The smallest minimum mass of a secure planet is 2.3~\mjup, while the smallest
minimum mass of a planet candidate is 1.1~\mjup. On the other hand, the largest
minimum mass among secure planets is 25~\mjup, while a few planet candidates
could have masses even more massive than that (but still rather uncertain 
because of long period orbits which still need to close). As mentioned 
earlier, we refer to all of these objects as planets (or planet candidates).

{\bfa Eight} of the 15 systems with a secure planet in our Lick sample have been 
published already:
\object{\iotadra~b}\ \citep{frink02}, \object{Pollux~b} \citep{reffert06,hatzes06},
\object{$\varepsilon$ Tau~b} \citep{sato07}, \object{11~Com~b} 
\citep{liu08}, \object{$\nu$~Oph~b and c} \citep{quirrenbach11,sato13},
\object{$\tau$~Gem~b} and \object{91~Aqr~b} \citep{mitchell13}, and
\object{$\eta$~Cet b and c} \citep{trifonov14}; we will publish
the remaining ones in the near future.

\section{Planet Occurrence Rate as a Function of Stellar Mass and Metallicity}

\subsection{Whole Sample}
In Fig.~\ref{fehmass}, we plot stellar mass as a function of metallicity
for all our stars in the Lick sample. We distinguish the 186 stars with
which we started in 1999/2000 (dots; `first sample') from the 196 ones which 
were added in 2004 with different selection criteria (crosses; `second sample'). 
The solid and dashed lines indicate the median masses per metallicity bin 
of the stars of the first and second sample, respectively. One can clearly see
that the stars of the second sample have higher masses than the stars
of the first sample, in particular at high metallicities. It was our
aim to study exoplanets around high mass stars when we put together 
the second sample, so this bias was introduced on purpose through our
selection criteria. 

\begin{table*}
\caption{Number of stars with planets (\np), with planet candidates (\nc)
and number of all stars (\ns) in our Lick sample for the bins in metallicity 
([Fe/H]$_{\mbox{\tiny bin}}$) and stellar mass ($M_{\mbox{\tiny *,bin}}$) shown in Fig.~\ref{fehmass}.
The corresponding planet frequencies for planets (\fpl) and for planets and planet 
candidates combined (\fpc) are also given, together with their 68.3\% confidence levels
derived from binomial statistics. 
}
\label{fehmasstable}
\centering
\begin{tabular}{cccccccc}
\hline\hline\noalign{\smallskip}
[Fe/H]$_{\mbox{\tiny bin}}$  & $M_{\mbox{\tiny *,bin}}$ & \np & \nc & \ns & \fpl & \fpc \\
{}[dex]                & [\msun]            &     &     &    & [\%]  & [\%]  \\
\hline\noalign{\smallskip}
--0.28\,\ldots\,--0.12 &  1.0\,\ldots\,1.8 &  0  &  5  & 41 & \phn0.0 $^{+4.3}_{-0.0}$ &    12.2 $^{+7.0}_{-3.4}$ \\[0.6ex]
                       &  1.8\,\ldots\,2.6 &  2  &  0  & 29 & \phn6.9 $^{+7.9}_{-2.3}$ & \phn6.9 $^{+7.9}_{-2.3}$ \\[0.6ex]
                       &  2.6\,\ldots\,3.4 &  0  &  0  & 21 & \phn0.0 $^{+8.0}_{-0.0}$ & \phn0.0 $^{+8.0}_{-0.0}$ \\[1.4ex]
--0.12\,\ldots\,+0.04  &  1.0\,\ldots\,1.8 &  2  &  0  & 29 & \phn6.9 $^{+7.9}_{-2.3}$ & \phn6.9 $^{+7.9}_{-2.3}$ \\[0.6ex]
                       &  1.8\,\ldots\,2.6 &  1  &  3  & 51 & \phn2.0 $^{+4.2}_{-0.6}$ & \phn7.8 $^{+5.5}_{-2.3}$ \\[0.6ex]
                       &  2.6\,\ldots\,3.4 &  0  &  1  & 45 & \phn0.0 $^{+3.9}_{-0.0}$ & \phn2.2 $^{+4.8}_{-0.7}$ \\[1.4ex]
 +0.04\,\ldots\,+0.20  &  1.0\,\ldots\,1.8 &  4  &  2  & 16 &    25.0 $^{+13.3}_{-7.7}$ &    37.5 $^{+12.9}_{-10.1}$ \\[0.6ex]
                       &  1.8\,\ldots\,2.6 &  4  &  1  & 24 &    16.7 $^{+10.2}_{-5.0}$ &    20.8 $^{+10.4}_{-5.9}$ \\[0.6ex]
                       &  2.6\,\ldots\,3.4 &  2  &  1  & 22 & \phn9.1 $^{+9.9}_{-3.1}$ &    13.6 $^{+10.5}_{-4.4}$ \\[0.3ex]
\hline\noalign{\smallskip}
\end{tabular}
\end{table*}

In order to study planet occurrence rate as a function of metallicity and mass,
we divided the area which is most populated in the metallicity/stellar mass 
plane into nine bins (three in each dimension); see Fig.~\ref{fehmass} 
and Table~\ref{fehmasstable}.
We then determined the fraction of stars with planets (filled circles) 
and planet candidates (open circles) in each of these bins. The resulting
planet occurrence rates (counting planets as well as planet candidates) are 
overplotted in Fig.~\ref{fehmass} in the center of each bin (percentage numbers). 

Table~\ref{fehmasstable} gives the number counts of planets, planet candidates
and stars in each bin, and the resulting planet occurrence rates counting
only secure planets (\fpl) and counting secure planets as well as planet
candidates (\fpc). 68.3\% confidence limits on planet occurrence rates are also
given; they were calculated based on binomial statistics following 
the Bayesian approach \citep{cameron11}.

From Fig.~\ref{fehmass} and Table~\ref{fehmasstable}, a rather clear pattern emerges:
planet occurrence rate increases strongly with metallicity in our sample, and it
decreases with stellar mass in the mass range which we probe here. 
Planet occurrence rate is highest (37.5 $^{+12.9}_{-10.1}$\%) in the bin with the 
highest metallicities and the lowest stellar masses, and it drops down to 0\% in the 
mass bin with the lowest metallicities and highest stellar masses. In between,
we observe intermediate values, so there seems to be a smooth transition
between the highest and lowest planet occurrence rates as a function of
metallicity and stellar mass.

This result does not change much when we reject the planet candidates and only
use the confirmed planets for the statistics, as one can see from a comparison
of the last two columns in Table~\ref{fehmasstable}. Of course, the overall planet
occurrence rate is lower if only confirmed planets are counted. The fraction
of stars with confirmed giant planets, \fpl, can be regarded as a lower value
for the true fraction of giant stars {\bfa harboring giant planets} in our sample, 
while the fraction of stars
which either harbor a confirmed {\bfa planet or a planet candidate}, \fpc, can be 
regarded as an upper value for that number. 

The correlations with metallicity and mass which we described above are present
irrespective of whether planet candidates are included or excluded.
However, if considering the distribution of planets and planet candidates separately, 
one notable difference emerges: we find a lot of planet candidates at rather small 
metallicity and mass. In our lowest mass and lowest metallicity bin, we only find 
planet candidates, but no planets. This trend continues to smaller metallicities. 
We do not know what the reason is for the different distribution of planets and
planet candidates with respect to metallicity, but it is certainly possible that some 
fraction of the planet candidates are not true planets.

We will now examine the correlation of planet occurrence rate with stellar metallicity
and stellar mass in more detail. 
For this analysis, we restrict ourselves to the planet population of the confirmed planets
only. We divided our Lick sample of stars into more metallicity and mass bins (namely, 20
bins over the full range of values shown in Fig.~\ref{fehmass}), and determined
planet occurrence rates in those bins. Since we are plagued by small number statistics with
such a large number of bins, we heavily smoothed the resulting planet occurrence rate
map using a sliding average over five bins, in order to reveal the general trends present
in the distribution of planets, but not the small, insignificant details. The resulting  
planet occurrence rate map is shown in Fig.~\ref{fehmassmap}. 

Planet occurrences rate is color coded as indicated by the bar on the right hand
side of the plot; the planet occurrence rate ranges from 0\% (the large area of the diagram
in light yellow, especially at small metallicity and high mass) up to 40\% for 
metallicities around 0.2 and masses around 2~\msun\ (black). The white areas of the diagram
correspond to metallicity/mass combinations for which we do not have any measurements, i.e.\
which are not covered by our Lick sample. 

\subsubsection{Fitted Dependencies}

In order to quantify the planet occurrence rate as a function of metallicity and stellar mass,
and in order to 
compare our results with the findings of others, we fitted our data to a model following
an exponential distribution in metallicity and a Gaussian distribution in stellar
mass, as suggested by Fig.~\ref{fehmassmap}. In addition, we also fitted our data
with a power law distribution in stellar mass \citep[see][]{johnson10}, which we
modified also to include an exponential cutoff so that it better describes the decrease 
in planet occurrence rate at higher masses. For comparison, we also fitted our
observations to a flat distribution in mass, keeping only the exponential distribution
in metallicity. 

The Gaussian plus exponential distribution
has the following form with parameters 
$C$, $\mu$, $\sigma$ and $\beta$:
\begin{equation}
\label{expgauss}
f(M_*,\mbox{[Fe/H]}) = C\,\exp\left({-\frac{1}{2}\left[\frac{M_*-\mu}{\sigma}\right]^2}\right)\,10^{\beta \mbox{\tiny [Fe/H]}} \qquad ,
\end{equation}
whereas the power law with cutoff plus exponential distribution 
is described by 
\begin{equation}
\label{exppower}
f(M_*,\mbox{[Fe/H]}) = C\,(M_*/\mbox{M}_{\odot})^{\alpha}\,\exp\left({-\frac{M_*}{M_0}}\right)\,10^{\beta \mbox{\tiny [Fe/H]}}   \qquad ,
\end{equation}
with parameters $C$, $\alpha$, $M_0$ and $\beta$.
{\bfa The power law plus exponential distribution with parameters $C$, $\alpha$ and $\beta$ 
as used by \citet{johnson10} has the form 
\begin{equation}
\label{eqjohnson}
f(M_*,\mbox{[Fe/H]}) = C\,(M_*/\mbox{M}_{\odot})^{\alpha}\,10^{\beta \mbox{\tiny [Fe/H]}}   \qquad .
\end{equation}
}
$f(M_*,\mbox{[Fe/H]})$ is the fraction of stars with giant planets, 
$M_*$ is the stellar mass and [Fe/H] is the stellar metallicity. 

\begin{figure}
\resizebox{\hsize}{!}{\includegraphics{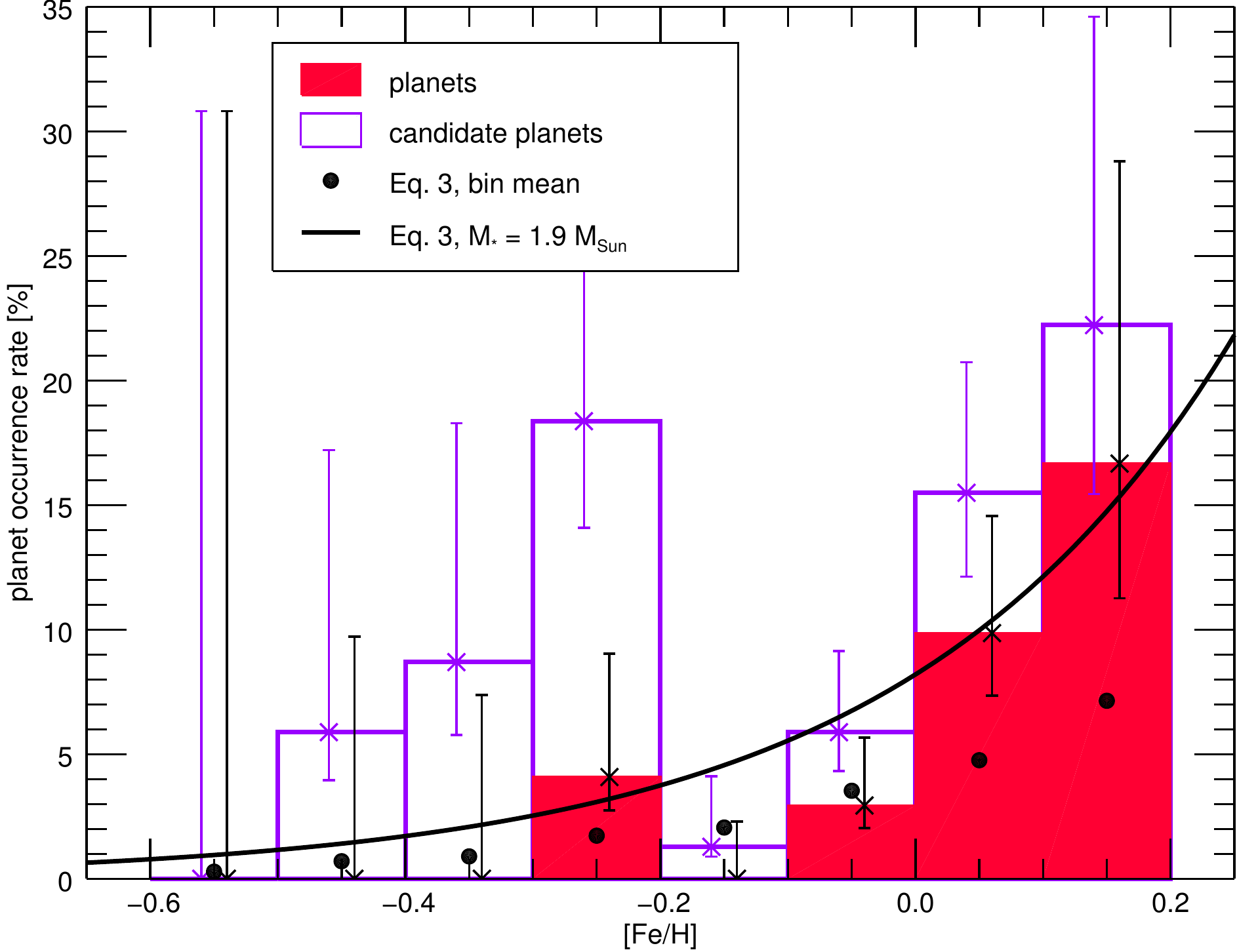}}
\caption{Planet-metallicity correlation observed in our Lick
sample, ignoring the effect of stellar mass on the planet occurrence rate. 
The filled histogram shows
secure planets, whereas the open histogram includes planet candidates
as well. Error bars are computed based on binomial statistics as explained
in the text. The solid line {\bfa illustrates the exponential fit to the
planet occurrence rate of secure planets as a function of metallicity, for 
a stellar mass of 1.9~\msun. The black dots also correspond
to the exponential fit, but here the individual mass distribution in each
bin has been taken into account.}
}
\label{feh_allmass}
\end{figure}

{\bfa We performed the fitting with two different methods: Levenberg-Marquardt least
squares minimization and Bayesian inference. The results were largely identical, so
we quote only the results of the Bayesian analysis in Table~\ref{fitparm}. The
disadvantage of the least squares minimization in our context is its inability to account 
for asymmetric error bars, as applicable for binomial population proportions. On the other
hand, the Bayesian technique has some shortcomings as well; we performed a simple
grid search for the maximum likelihood, which is computationally expensive, and the 
choice of priors or meaningful parameter intervals is rather arbitrary. 

An advantage of Bayesian inference for the given analysis is the possibility
to account for a probability distribution in mass, rather than assuming one
fixed value with an error bar. We constructed these probability distributions 
by adding two Gaussians (where applicable), which were centered on the two different mass 
estimates which we derived for most of our stars (Table~\ref{allstars}). 
The formal errors of the masses correspond to the widths of the Gaussians, and
the relative heights were given by the probabilities derived for the red giant branch 
and the horizontal branch solution, respectively. We thus ended up with bimodal 
distributions for the masses, although in many cases the two peaks in the mass probability
distribution are located close to each other. 

Our strategy for applying the principles of Bayesian inference closely followed
the one outlined in \citet{johnson10}. Specifically, we would like to stress that we
fitted the various planet occurrence models to each star individually, not to binned 
or smoothed data. We chose uniform priors for all parameters, since we did not want
to bias the result in any way. We ensured that all prior ranges included the peak
in the maximum likelihood as well as the 68.3\% confidence regions of the parameters.

The result of the Bayesian fitting is summarized in Table~\ref{fitparm}. 
The first line gives the limits of the uniform prior ranges. The parameters given
for each model correspond to the peak in the likelihood function; the 68.3\% confidence
intervals are quoted in the second line for each fitted model. Most posterior probability
distribution functions are slightly asymmetric; the quoted confidence intervals thus
correspond to the shortest intervals with a 68.3\% chance for containing the correct parameter 
values. The Bayes factor $B$ is the ratio of the evidence of a given model divided by our
best fitting model described by Eq.~\ref{expgauss}. Bayes factors between 1 and 3.2
are `not worth more than a bare mention', whereas Bayes factors between 10 and 100
provide `strong evidence' against the model being tested \citep{kass95}. 

\begin{figure}
\resizebox{\hsize}{!}{\includegraphics{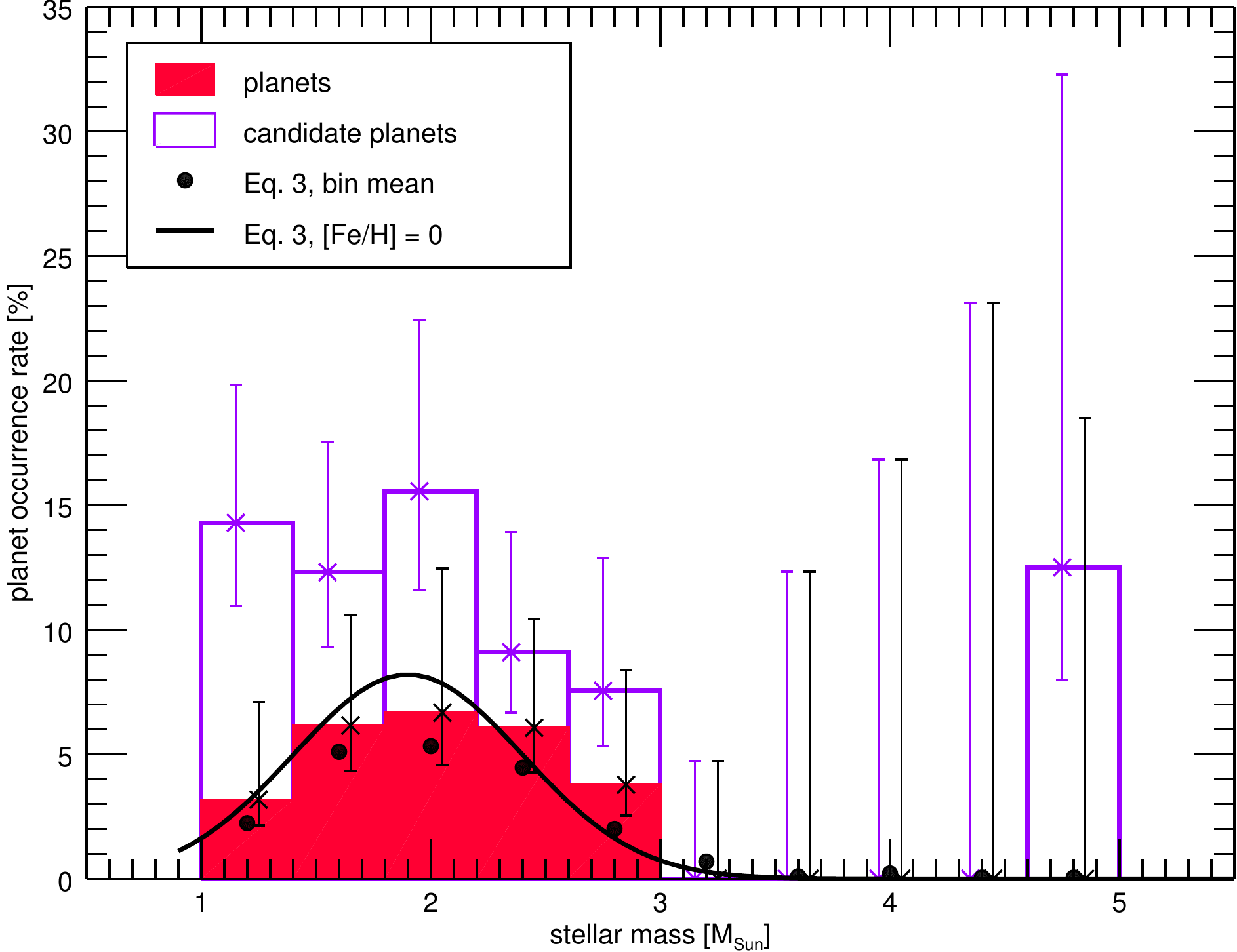}}
\caption{Planet occurrence rate as a function of stellar mass in our Lick sample,
{\bfa ignoring the effect of stellar metallicity}.
See caption of Fig.~\ref{feh_allmass} for the explanation of histogram data and error
bars. {\bfa The solid line denotes our best fit to the mass dependence of the giant planet
occurrence rate computed for zero metallicity. The black dots correspond to the same
model, but the true metallicity distribution within each bin has been taken into account.}
}
\label{allfeh_mass}
\end{figure}

\begin{table*}
\caption{{\bfa Derived parameters, 68.3\% confidence regions and Bayes factors $B$
obtained via Bayesian} fitting of observed planet occurrence
rates as a function of stellar mass and metallicity; 
see Eqs.~\ref{expgauss}, \ref{exppower} and \ref{eqjohnson}.
Parameter values from other studies are given for comparison.}
\label{fitparm}
\centering
\begin{tabular}{lcccccccc}
\hline\hline\noalign{\smallskip}
  & \raisebox{-1.5ex}[1.5ex]{$\alpha$} & \raisebox{-1.5ex}[1.5ex]{$\beta$} 
  & \raisebox{-1.5ex}[1.5ex]{$C$} & $\mu$ & $\sigma$ & $M_0$ & \raisebox{-1.5ex}[1.5ex]{B} & mass range \\[-1.4ex]
  &          &         &     & [\msun] & [\msun] & [\msun] & & [\msun] \\[0.2ex]
\hline\noalign{\smallskip}
\multicolumn{9}{c}{new Bayesian fits} \\
\hline\noalign{\smallskip}
uniform prior limits & --1.5\dots+1.5 & 0.0\ldots3.0 & 0.01\ldots0.30   & 1.0\ldots3.0 & 0.1\ldots2.0 & 0.5\ldots2.0 \\
\hline\noalign{\smallskip}
Eq.~\ref{expgauss}   & \ldots         & 1.7       & 0.082         & 1.9       & 0.5       & \ldots    & 1.0     & 1.0 \ldots 5.0 \\
                     & \ldots         & 1.3 - 2.0 & 0.056 - 0.122 & 1.4 - 2.0 & 0.3 - 1.0 & \ldots    & (fixed) &     \ldots     \\[0.8ex]
Eq.~\ref{exppower}   & 1.0            & 1.8       & 0.223         & \ldots    & \ldots    & 0.7       & 1.3     & 1.0 \ldots 5.0 \\
                     & (fixed)        & 1.3 - 2.0 & 0.111 - 0.432 & \ldots    & \ldots    & 0.6 - 1.0 & \ldots  &     \ldots     \\[0.8ex]
Eq.~\ref{eqjohnson}  & --0.3          & 1.6       & 0.089         & \ldots    & \ldots    & \ldots    & 8.4     & 1.0 \ldots 5.0 \\
                     & --0.6 - 0.0    & 1.2 - 1.9 & 0.059 - 0.124 & \ldots    & \ldots    & \ldots    & \ldots  &     \ldots     \\[0.8ex]
flat                 & \ldots         & 1.8       & 0.044         & \ldots    & \ldots    & \ldots    & 76      & 1.0 \ldots 5.0 \\
                     & \ldots         & 1.3 - 2.1 & 0.031 - 0.060 & \ldots    & \ldots    & \ldots    & \dots   &     \ldots     \\
\hline\noalign{\smallskip}
\multicolumn{9}{c}{other investigations} \\
\hline\noalign{\smallskip}
\cite{johnson10}     & $1.0 \pm 0.3$  & $1.2 \pm 0.2$   & $0.07  \pm 0.01$  & \ldots          &  \ldots         & \ldots          & & 0.2 \ldots 2.0\\
\cite{udry07}        & \ldots         & $2.04$          & $0.0301$          & \ldots          &  \ldots         & \ldots          & & 0.7 \ldots 1.4\\
\cite{fischer05}     & \ldots         & $2.0$           & $0.03$            & \ldots          &  \ldots         & \ldots          & & 0.7 \ldots 1.5\\
\hline\noalign{\smallskip}
\end{tabular}
\end{table*}

The Gaussian distribution in mass
(Eq.~\ref{expgauss}) and the power law with cutoff (Eq.~\ref{exppower}) fit the data about 
equally well. Our best fitting model is the one with the Gaussian distribution in mass; 
we obtain $C = 0.082^{+0.040}_{-0.026}$, $\beta = 1.7^{+0.3}_{-0.4}$, 
$\mu = 1.9^{+0.1}_{-0.5}$~\msun\ and $\sigma = 0.5^{+0.5}_{-0.2}$~\msun. 
The parameter $\mu$ gives the stellar mass with the highest probability for the presence
of a giant planet. Furthermore, according to this model,
the planet occurrence rate has dropped to half of its peak value at masses of
1.2~\msun\ and 2.6~\msun, respectively. The uncertainty in these masses is of
the order of 0.5~\msun\ (combined error in both $\mu$ and $\sigma$).

A flat distribution in mass or the simple power law from 
\citet{johnson10} perform much worse; the evidence against those latter two models is `strong' 
\citep{kass95},
so that they can be rejected. The power law exponent $\alpha$ in the model from \citet{johnson10}
is even negative when fitted for with our data, suggesting that the decrease of planet
occurrence rate at higher masses dominates over its increase observed for smaller masses. 
This is why we fixed $\alpha$ to 1, the value obtained by \citet{johnson10},
when modeling our observations according to Eq.~\ref{exppower}. We also tried to fit for
$\alpha$ here, but did not succeed; the parameter is not well constrained by our data,
and is correlated with $M_0$. 
}

The value that we find for $\beta$ is fully
consistent with other studies on the planet-metallicity correlation, in particular
with \cite{fischer05} and \cite{udry07} (Table~\ref{fitparm}).
For the mass range from 0.2--2\,\msun, \cite{johnson10} found a
positive correlation between planet occurrence rate and stellar mass.
It seems that this correlation turns into an anticorrelation for masses
larger than about {\bfa 1.9\,\msun.}

In Fig.~\ref{feh_allmass} we show the planet occurrence rate as
a function of metallicity, and in Fig.~\ref{allfeh_mass} we show the planet occurrence
rate as a function of stellar mass, respectively. Separate histograms are plotted 
for either only the secure planets or including planet candidates. 
{\bfa The solid line and the black dots indicate the fit which was obtained
above from applying Eq.\ref{expgauss} to the {\bfa individual data} (secure planets
only), not the histogram data shown in the plots which either ignores
the dependence of planet occurrence rate on mass (Fig.~\ref{feh_allmass}) or 
on metallicity (Fig.~\ref{allfeh_mass}). It consists of 
an exponential for the dependence on metallicity (Fig.~\ref{feh_allmass})
and a Gaussian for the dependence on stellar mass (Fig.~\ref{allfeh_mass}).}

Ignoring the effect of stellar mass (Fig.~\ref{feh_allmass}, histogram data), 
the planet-metallicity correlation {\bfa seems even}
steeper than compared to the fit that takes both metallicity and stellar
mass into account (Fig.~\ref{feh_allmass}, black dots). {\bfa It is also
evident from Fig.~\ref{feh_allmass} that the planet candidates are distributed 
differently with respect to stellar metallicity than the secure planets; no clear
planet-metallicity correlation is seen for the planet candidates alone.}

{\bfa The distribution of secure planets and planet candidates, respectively, with
respect to stellar mass (Fig.~\ref{allfeh_mass}) is rather similar to each other;
we do not see any differences in the two distributions such as those observed 
in the planet-metallicity correlation of the two planet samples.}

\begin{figure}
\resizebox{4.4cm}{!}{\includegraphics{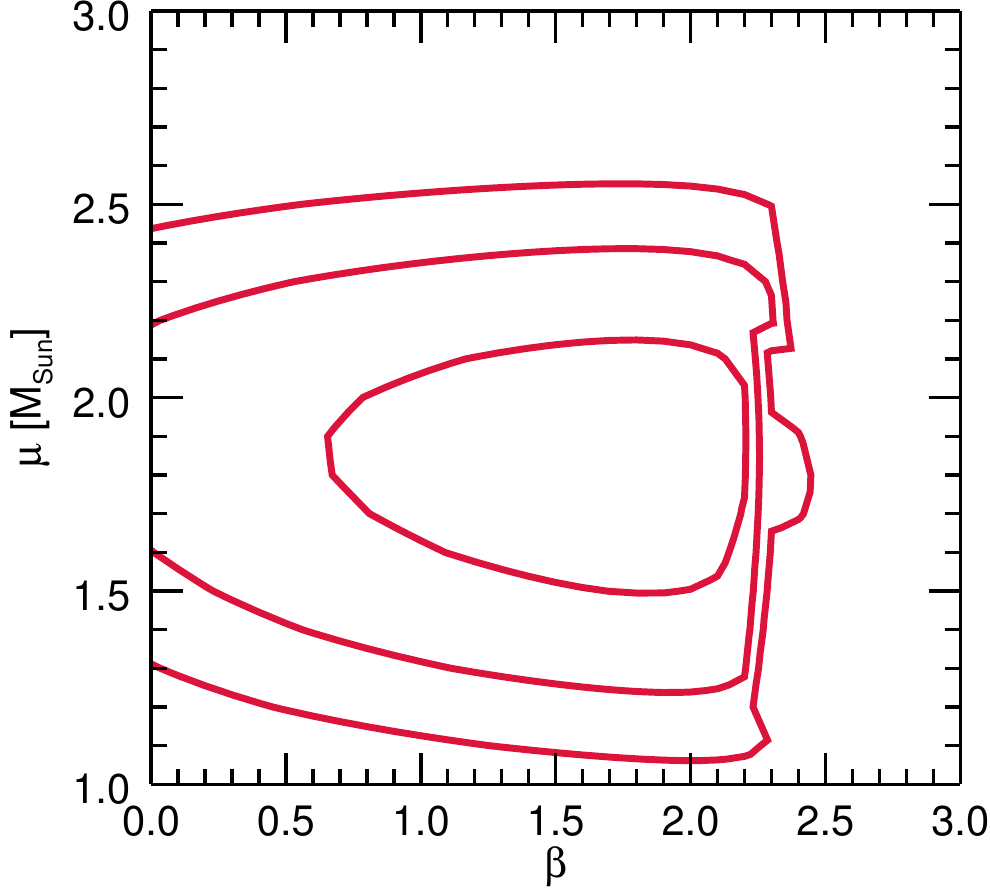}}\hfill
\resizebox{4.4cm}{!}{\includegraphics{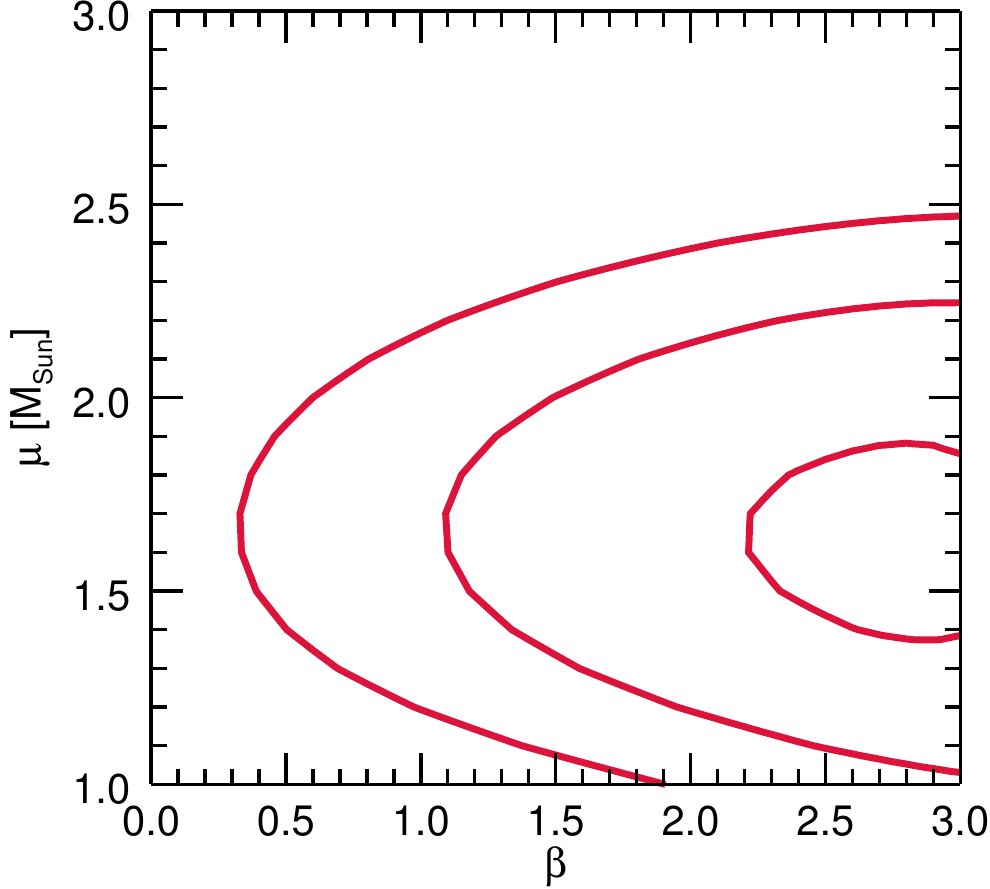}}
\caption{{\bfa Contours of equal Bayesian likelihood as a function of parameters
$\beta$ and $\mu$ for our best-fitting model (Eq.~\ref{expgauss}). The panel on
the left corresponds to the full Lick sample, while the panel on the right
corresponds to the subsample. In both cases no correlations between the
parameters $\beta$ (related to metallicity) and $\mu$ (related to mass) can be 
seen, indicating that the giant planet occurrence rate indeed depends on both,
stellar mass and metallicity. }
}
\label{contour}
\end{figure}

{\bfa Fig.~\ref{contour} shows contours of equal Bayesian likelihood as a function of
parameters $\beta$ and $\mu$ of our best-fitting model (Eq.~\ref{expgauss}). 
There is no correlation
between the two parameters characterizing the dependence of giant planet
occurrence rate on stellar mass and metallicity, respectively. This indicates that
the giant planet occurrence rate depends on both parameters directly,
independently of each other, and is not
due e.g.\ to the uneven distribution of stars in the mass-metallicity plane.}

The most striking result of our analysis is clearly the sharp decrease in planet occurrence 
rate for stars with masses higher than 2.5 to 3~\msun. 
The highest mass of a planet bearing star in our sample is 2.7~\msun. We do not
find any confirmed planets around stars with larger masses, although there are
119 such stars in our sample (83 of those stars are in the mass range from 2.7 to 3.5~\msun). 
This corresponds to a giant planet occurrence rate of $<1.6$\% at 68.3\% confidence,
and $<5.6$\% at 99.73\% confidence in the mass interval from 2.7 to 5.0~\msun.  
We will discuss the dependence of giant planet occurrence
rate on stellar mass further in Section~\ref{discussion}.

\subsection{Subsamples}
\label{subsample}

In the previous section we have taken our full sample as a reference, 
assuming that we would be able to detect giant planets with periods up to a 
few years around any star in our sample. Here we will show that even
when cutting down our sample to ensure uniform detectability of planets,
the results of the previous section still hold. 

There are three parameters that influence {\bfa our capability to detect} a 
given planet around a particular star in our sample: stellar mass, 
intrinsic stellar jitter, and number of observations. 
Observing time span does not matter in our context, because all stars have been 
observed for at least six years. With the exception of the period of the outer companion
in the resonant double brown dwarf system $\nu$~Oph, the periods for confirmed
planets in our sample range from 0.5 to 2.3~years.

The radial velocity semi-amplitude $K_1$, which is imposed by a planet 
with given mass $m_2$, period $P$ and eccentricity $e$ on a star of mass $M_*$,
is given by
\begin{equation}
K_1 = \left(\frac{2 \pi G}{P}\right)^{1/3}\frac{1}{\sqrt{1-e^2}}\,\,\frac{m_2\sin i}{\left(M_* + m_2\right)^{2/3}}\qquad,
\label{eq:amplitude}
\end{equation}
where {\bfa $i$ is the inclination} and $G$ is the gravitational constant. 

We consider a given planet around a given star detectable in our survey if
the radial velocity semi-amplitude that it would generate is larger than
the intrinsic stellar jitter (any orbital motion is subtracted from the
observed velocities before the intrinsic stellar jitter is derived
from the standard deviation of the radial velocities). 
Furthermore, we require at
least 12 observations for each star, in order to allow for the detection of
the planetary signal in the presence of stellar jitter. 
The median intrinsic stellar jitter in our sample is 22~\ms, while the RV 
semi-amplitude $K_1$ of a planet with a mass of 2.3~\mjup\ (smallest
mass among our confirmed planets), period of 2.3~years (largest period
among our confirmed planets) around a 2~\msun\ star (typical stellar mass) 
is 31~\ms. 

{\bfa For comparison, the detection capability threshold used by \citet{fischer05}
corresponds to an RV amplitude of at least 30\,\ms, a period of less than 4~years,
and at least 10 observations per star, very similar to our requirements.
The subgiant planet hosts analyzed statistically by \citet{johnson10} also have
Doppler signals which are typically larger than 20--30~\ms, so that the 
detection threshold between the various surveys is not that different; we only
miss a few planets with very small RV amplitudes in our Lick giant star
survey in comparison with other surveys. 

However, we caution that the resulting planets to which we are sensitive will be
more massive, for two reasons: (1) Since the radial velocity signal depends on
stellar mass (Eq.\ref{eq:amplitude}), and our stars are more
massive than those in typical main-sequence Doppler samples (average mass of about 
2\,\msun\ for giant stars as opposed to 1\,\msun\ for main-sequence stars), 
the resulting planet masses in our sample are larger by 
a factor of about 1.6. The same is true, though with even smaller differences
in resulting planet masses, for subgiants (typical stellar mass 1.5\,\msun).  
(2) The distribution of orbital parameters of planets 
found around giant stars and those found around main-sequence or subgiant
stars differ, most notably in period. While a few hot Jupiters are known
around subgiant stars \citep{johnson06,johnson10b}, such short-period planets 
are completely absent from giant star samples;
they are either engulfed by the star or move further out as the radius of the 
stellar host increases \citep{sackmann93,villaver09}. In fact, roughly half of our
stars are on the horizontal branch already (as opposed to the red giant branch), and
have thus experienced a short phase with a very expanded shell at the end of the 
red giant branch, potentially altering the observable planets' orbital parameters. 
Larger period however means smaller RV signal, which can only be compensated by
a larger planet mass.}

Our goal is to assemble a subsample that is more 
uniform in planet detectability than our full Lick sample, in order to 
eliminate any {\bfa potential} biases in the planet occurrence rate as a function
of metallicity and stellar mass that result from reduced planet detection
capabilities around stars of high mass, large jitter or poor observing 
history. For each star in the full Lick sample, we check whether a
planet with a minimum mass of 2.3~\mjup\ (which corresponds to the smallest
planet mass among our confirmed planets) and a period of 2.3~years (which
corresponds to the longest period among our confirmed planets, with the
exception of the outer companion to $\nu$~Oph) would be detectable 
according to the definition above. 

We end up with 207 (out of 373) stars that fulfill
the detectability criteria and which thus constitute our subsample. 
{\bfa Among the rejected stars are three planet-bearing stars and seven
additional stars hosting planet candidates. While we managed to 
detect a (massive or small period) planet around these stars, it means we would not 
have been able to detect the small planet which we defined as reference 
for detectability.}

The subsample still contains 50 out of the 119 stars with stellar masses
larger than 2.7\,\msun\ from our original sample. Thus, the fraction of
high mass stars in the subsample (24\%) is a bit smaller than in the full sample
(32\%), but not by as much as one might expect. 
The reason is that the more massive
stars in our sample have less than average jitter (due to our original
selection criteria), which makes up for the reduced planet detection
capability around a star of larger mass. 

{\bfa In Fig.~\ref{sample_comp} we show a comparison of planet occurrence rates 
in the full sample and the subsample for the same bins as in
Table~\ref{fehmasstable}. One can clearly see that the planet occurrence rates
in the various stellar mass and metallicity bins are very similar between the full
and the restricted samples.
The small differences are not significant if the errors are taken into
account. Errors vary greatly across bins, but are typically of the order of
5-10\% for the full sample (see Table~\ref{fehmasstable}), and not very 
different for the subsample (typically of the order of 1\% larger). 

Furthermore, one can clearly discern the two bins where planet-bearing stars
were rejected for the subsample; these are those where the planet occurrence
rate drops. The only two planet-bearing stars in the lowest metallicity and
medium mass bin were removed, as well as one planet-bearing star in the
highest metallicity highest mass bin.
For the remaining bins, planet occurrence rates either remain at zero or
increase slightly, reflecting the removal of a few stars without planets in
those bins. 

The overall pattern in the dependence of planet occurrence rate with mass and
metallicity which we observed for the full sample is still clearly visible
for the subsample and does not appear to have changed significantly.
We fitted the planet occurrence rate for the subsample 
as a function of stellar mass and metallicity to the same models as for the
full sample. Unfortunately, the Bayesian fitting did not work as well for
the subsample as for the full sample, presumably because increasingly large
parts of the mass-metallicity plane are not sampled well, and the overall number
of stars is small. This means that we cannot assign reliable confidence limits
to the derived parameters; we still managed to derive best-fit values for all
parameters from a clear maximum in the likelihood function. 
Least-squares minimization worked well though, and was consistent with the
results from Bayesian fitting. Yet, for consistency with the last section, we quote
the results from Bayesian fitting for the subsample as well. 

\begin{figure}
\resizebox{\hsize}{!}{\includegraphics{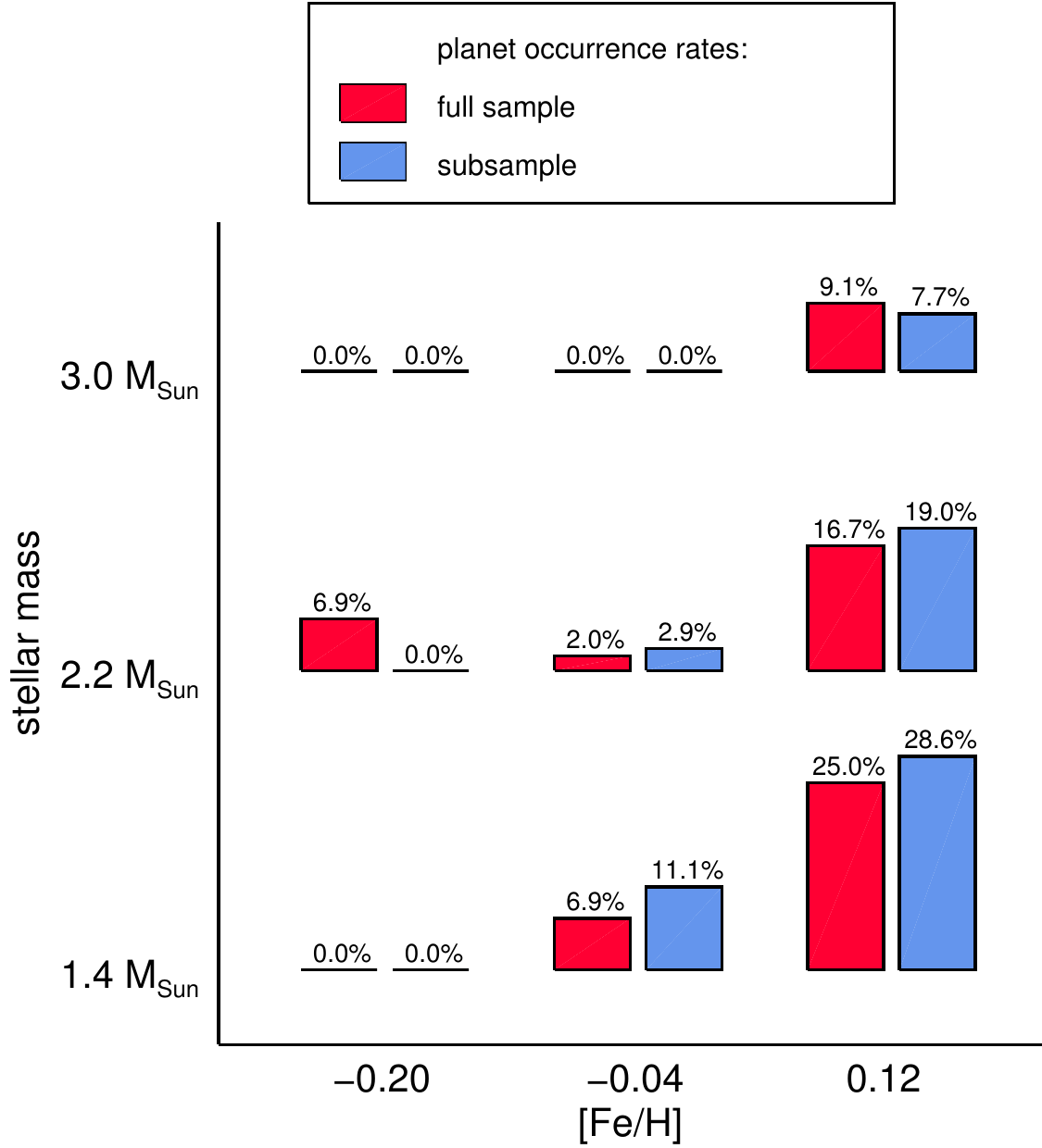}}
\caption{{\bfa Comparison of the planet occurrence rates for our full sample and the 
subsample which ensures uniform detection capability (ensuring detectability of a planet of
at least 2.3~\mjup\ with a period up to 2.3~years). The planet occurrence
rates in both sample are very similar in the various stellar mass and metallicity
bins; the small differences are not significant.}
}
\label{sample_comp}
\end{figure}

Just as for the full sample, the model consisting of an exponential
distribution in metallicity and a Gaussian distribution in mass (Eq.~\ref{expgauss})
performed best. We obtained the following parameters: 
$C = 0.079$, $\beta = 2.8$, $\mu = 1.6$\,\msun, and $\sigma = 0.6$\,\msun. 
These parameters are very similar to those derived for the full sample, with
the exception of the parameter $\beta$, which indicates a steeper 
planet-metallicity correlation than for the full sample. However, considering that
the errors for the subsample are certainly somewhat larger than for the
full sample, the results are fully consistent with each other, and the 
differences are not significant. 

The fitting of a four parameter model to the subsample data is clearly
stretching the statistical analysis to the limits; after all, the subsample 
data just consist of
12 planets around 207 stars, distributed over mass and metallicity space. 
Clearly, a much larger sample with uniform detection capability with respect
to planets would be desirable, but will not be available in the near future.
Nevertheless, the subsample data clearly support our most important finding,
namely that the giant planet occurrence rate decreases sharply for stellar masses
larger than 2.5--3.0\,\msun. In the subsample, we still have 50 stars with
masses larger than 2.7\,\msun, around which no planet has been found.
This corresponds to a giant planet occurrence rate smaller than 3.5\% at 68.3\%
confidence, significantly different from the giant planet occurrence rates 
for lower mass and higher metallicity bins.
}

From the comparison of results for the full sample and the subsample
we conclude that no obvious biases are present in the analysis of our full sample.
We confirm a strong planet-metallicity correlation based on our subsample analysis,
and a peak in planet occurrence rate for stellar masses somewhere in the interval
between 1.5 and 2\,\msun, with a strong decline in giant planet occurrence rate
for stars with masses larger than 3\,\msun.

\section{Discussion}
\label{discussion}

\subsection{Correlation of Giant Planet Occurrence Rate with Stellar Metallicity}

We find a strong planet-metallicity correlation in our Lick sample of G and K giants,
with about the same power law exponent (to within the errors) as observed for main-sequence stars
by \cite{fischer05} and \cite{udry07}; see Table~\ref{fitparm}).
\cite{johnson10} also observed a positive planet-metallicity correlation in a sample of
subgiant stars, yet with a smaller power law exponent (possibly because of the simultaneous
fitting for the correlation with stellar mass and the distribution of stars in the
mass-metallicity plane).

{\bfa The parameter $\beta$ which describes the planet-metallicity
relation is remarkably constant across models, irrespective of the model used to describe
the mass dependence of the giant planet occurrence rate. This indicates already that we
are really dealing with two separate effects here, and planet occurrence rate does indeed
depend on both, stellar mass and stellar metallicity.}

However, our finding is in sharp contrast to the results of \cite{pasquini07}
and \cite{takeda08}, who did not find the same correlation of planet occurrence 
rate with metallicity as the one observed for main-sequence stars 
in their giant star samples. One reason could be that their
selection criteria for confirmed planets were not as strict as ours. In fact, our
planet candidates do not show a strong planet-metallicity correlation; many 
planet candidates are found around lower-metallicity stars. It is indeed
notoriously difficult to distinguish giant stars harboring true planets from
giant stars which show (multi-)periodic radial velocity variations not due to planets
(see Section~\ref{sample}). We only used secure planets when deriving
our planet-metallicity correlation, which is essential. Since the secure planet
and the planet candidate samples show a different distribution with respect to 
metallicity, we believe that our sample of planet candidates is a true mixture of 
stars with true planets and stars without any planets. 

In fact, this same 
argument can also explain the results from \citet{maldonado13}, who noted a 
difference in planet occurrence rate and metallicity between low-mass and 
high-mass giant stars. If we divide
our sample of secure planet and planet candidate hosts at a stellar mass of 1.5\,\msun\
just as they did,
we also do not observe a strong planet-metallicity correlation for the lower mass
sample, but a strong planet-metallicity correlation for the higher mass
sample (see Fig~\ref{fehmass}). The reason for this is that the planet candidates are 
preferentially found
at lower metallicity and also at lower mass and thus pollute the strong 
planet-metallicity correlation of the secure planets. The different findings for
lower and higher mass giant stars could thus simply be a consequence of a polluted
sample of planet hosts with non-planet bearing stars. 

We conclude that there are strong indications for a planet-metallicity correlation 
among giant star planet hosts which matches the observed planet-metallicity 
correlation for main-sequence planet hosts.

\subsection{Anticorrelation of Giant Planet Occurrence Rate with Stellar Mass}

The most striking result of our analysis of giant planet occurrence rate as a function
of stellar metallicity and stellar mass is clearly the steep decline of planet 
occurrence rate for masses larger than 2.5 to 3\,\msun. Indeed we do not find any
confirmed planet around a star with a mass higher than 2.7\,\msun, although there
are 113 such stars in the sample. Thus, the planet occurrence rate for stars with masses
in the range 2.7--5\,\msun\ is consistent with zero ($0.0^{+1.6}_{-0.0}$\%). 

Our Lick K giant survey is the first Doppler survey that covers the mass range 
between 2 and 5\,\msun; traditional Doppler surveys have
looked primarily at late-type main-sequences stars with typical masses between
0.7 and at most 1.5\,\msun\ \citep{fischer05,udry07}.
A notable exception is the Doppler survey of subgiants
by Johnson \citep{johnson10}, which covers stellar masses up to about 2\,\msun. 
Interestingly, \cite{johnson10} found that {\bfa the} planet occurrence rate increases
with mass up to masses of about 2\,\msun.

This is also the case in the Lick K giant survey: stars with masses of about 2\,\msun\
have the highest planet occurrence rates observed, significantly higher than those
for stars of about 1\,\msun. However, this trend reverses for even higher stellar 
masses, beyond about 2.5--3\,\msun. 

\citet{burkert07} have identified a lack of massive planets at intermediate
orbital distances for main-sequence stars in the mass range from 1.2--1.5\,\msun\ 
as compared to planets found around stars with a mass range from about 0.7--1.2\,\msun.
These differences in planet properties occur at smaller masses than relevant
here, yet this might be a first indication that higher stellar masses 
adversely affect planet formation.
\citet{burkert07} attributed the observed differences in planet properties to a
shorter disk depletion timescale for more massive stars, which could be caused
by either a smaller disk size or stronger extreme ultraviolet flux as compared to 
those around lower mass stars. 

\cite{ida05} and \cite{kennedy08a} both considered giant planet formation around
stars of various masses using a semi-analytic formation model. The details of the
models differ significantly, yet both predict an increase
in the giant planet formation rate with mass over the range from about 0.5 to
1.5\,\msun\ \citep{ida05} and from about 0.5 to 3.0\,\msun\ \citep{kennedy08a}, 
respectively \citep[see Fig.~7 of][]{kennedy08a}. After a sharp maximum, the giant
planet formation rate strongly decreases with stellar mass in both models.
Qualitatively, this is exactly what we observe, at a mass which falls
exactly in the mass range predicted by those two models.

The increase in planet occurrence rate with stellar mass can be
interpreted as being the result of the larger disk masses of
more massive stars. Neptune mass planets are frequent
around lower mass stars according to those models, consistent with
observations, but giant planets form only rarely around
lower-mass stars \citep{johnson07}. Higher disk masses increase the chance
of forming a planet in general and a giant planet in particular, and thus
the overall planet occurrence rate correlates with stellar mass.

The reason why this positive correlation between planet occurrence rate and 
stellar mass breaks down at masses larger than about 2.5--3\,\msun\ can
be understood as follows. As the mass of the star increases, the snow line,
at or beyond which most planets are thought to form, is located increasingly further
out ($a_{\mbox{\tiny snow}}\propto M_*^{\alpha}$, with $\alpha$ between 1 and 2
\citep{ida08, kennedy08b}. At increasing distances from the star, gas densities
and Kepler velocities are smaller, which both slows down growth rates. 
As a result, the migration time scale $\tau_{\mbox{\tiny mig}}$ is also longer 
\citep[$\tau_{\mbox{\tiny mig}} \propto M_*^{1.5}$,][]{kennedy08b}, while stellar
disk dispersal is faster \citep{kennedy09}. 
Thus, the slower growth rate,
increased migration time scale and short lifetime of the protostellar disk
might prevent stars with masses larger than $2.5-3$\,\msun\ from forming
giant planets that would be observable at semi-major axes of a few AU today, 
as probed by our survey.

\subsection{Effect of Stellar Evolution?}

Little is known about the fate of planets as the host star evolves from the
main sequence up the giant branch. Some calculations have been carried out
specifically for the solar system, with specific consideration of the 
question whether or not the Earth will eventually be engulfed by the Sun during
its post-main sequence evolution.

In the model by \citet{sackmann93} the Earth can evade being swallowed, 
whereas the model by \citet{schroeder08} predicts that the Earth will
be engulfed by the Sun before it reaches the tip of the red giant branch.
All studies agree, however, on an increase of Earth's orbital distance
during the early post-main sequence evolution of the Sun, which is 
due to conservation of angular momentum as the Sun loses mass; mass loss 
dominates over tidal interactions between the Earth and the Sun
as well as over dynamical drag, which both tend to shrink Earth's orbit. 
The disagreement on the final fate of the Earth mainly comes from different 
assumptions about mass loss, to which the models are rather sensitive. 

\cite{sato08a} have discussed possible planet engulfment during post-main sequence
evolution in order to
explain the observed lack of close-in planets for intermediate mass stars, while
\cite{villaver09} and \cite{kunitomo11} have derived the orbital evolution
of planets during the post-main sequence evolution of stars {\bfa more massive} 
than the Sun.
Their model calculations included changes in the orbital distance
of a planet induced by stellar mass loss and stellar tides, through conservation
of angular momentum; an essential part of the model calculations are
stellar evolutionary models. \cite{villaver09} also included frictional and gravitational 
drag forces, but since their effect was found to be negligible those effects were 
not included by \cite{kunitomo11}. The result of the model calculations are the 
time-dependent 
orbital distances of planets during the stars' red giant branch and horizontal 
branch phases of stellar evolution, from which the 
minimum orbital distance of a planet required to avoid engulfment
by the star (critical initial semi-major axis or survival limit) can be derived. 

In summary, the orbital distances of the inner planets become smaller
(eventually leading to engulfment), while
the orbital distances of the outer planets become larger during the red giant
and horizontal branch stages of the parent star.
The largest changes in the
orbital distances occur at the tip of the red giant branch, when the stellar
radius is largest and tidal effects become much more important. The 
critical initial semi-major axis of a planet to avoid engulfment by the parent 
star is strongly dependent on stellar mass, and to a lesser extent on 
planet mass and stellar metallicity, with a sharp transition at stellar masses 
of around 2~\msun\ which is due to the presence of the helium flash leading to an 
increase in 
stellar radius for stars with masses smaller than about 2\,\msun\ and the
absence of such an effect for stars with masses larger than about 2\,\msun\
\citep{kunitomo11}. The detailed predictions
for the critical semi-major axis for survival are quite different between
the models from \cite{villaver09} and \cite{kunitomo11}, presumably due to
differences in their assumptions about stellar evolution.

These studies illustrate that the properties of any planetary system will 
be subject to profound changes during the post-main sequence evolution of
its host star. It is thus possible that the properties of the planet population 
which we observe in our Lick giant star sample are not the same as 
they were while the host stars were still on the main sequence. Depending on the
age of the stars, significant orbital evolution could already have taken place.
Age determination of the giant stars is rather difficult, but we estimated 
the probabilities for each star to either be on the red giant branch or on 
the horizontal branch, respectively (given in Table~\ref{allstars} in the appendix). 
We find that the parent stars of 12 out of 15 confirmed planets (80\%) are more 
likely on the horizontal branch than on the red giant branch; for the 
planet candidates, only 8 out of 20 parent stars (40\%) are on the horizontal branch.
For comparison, 41\% of the stars in the whole Lick sample are on the horizontal
branch, 56\% are on the red giant branch; for the remaining 3\% we could not
determine an evolutionary state. 
Thus the fraction of stars with confirmed planets that are already on the 
horizontal branch is relatively large, and some orbital evolution could 
certainly have taken place already.

\cite{villaver09} and \cite{kunitomo11} compare their critical
survival orbital distances with the orbital distances of planets detected around
stars more massive than 1.5\,\msun, and find that all observed planets orbit
at orbital distances large enough to avoid engulfment during the red giant and
horizontal branch stellar evolutionary phases, and especially the phase at 
the tip of the red giant branch when the stellar radius is largest. 
Our confirmed planets
have orbital distances larger than the survival limit for the models
of \cite{kunitomo11} only, while it looks as if some of their orbits could
be in conflict with the predictions by \cite{villaver09}. However, for a
definitive conclusion one would have to compute specific limits for every
system separately, taking its orbital characteristics and masses into account.

Most importantly, \cite{kunitomo11} find that for stars more massive than
about 2.5\,\msun, there is a large gap between the critical orbital distance
and the orbital distance distribution found among observed planets. Thus, it
looks as if another mechanism besides engulfment plays a major role in
shaping the observed semi-major axis distribution of giant planets around 
intermediate mass stars, preventing orbits too close in. \cite{kunitomo11}
conclude that most likely giant planet formation is hindered for stars more
massive than about 2.5\,\msun. 

Due to the large gap between the predictions for the survival limit and
the observed distribution of orbital distances, planet engulfment does not play 
a major role for the interpretation of the observed planet occurrence rate 
as a function of mass for the Lick sample. It is very well possible that
some planets around stars in our sample might have been engulfed already,
but this could not explain why we do not find any giant planets around
stars more massive than 2.5--3.0\,\msun, in particular since the survival
line is much closer in for stars more massive than about 2\,\msun\ than
for less massive stars. Most likely, giant planets around stars more massive
than 2.5--3\,\msun\ are not being formed at separations of a few AU in the first 
place, rather than being engulfed or kicked out during stellar evolution later on.

\section{Summary}

We have analyzed the distribution of planet-bearing stars in our 
Lick sample of 373 G and K giants with respect to metallicity
and stellar mass. Altogether, we find secure planets around 15 stars
and planet candidates around 20 stars, from a Doppler survey running
over twelve years. We are sensitive towards giant planets; minimum planet
masses for the secure planets range from 2.3 to 25\,\mjup.
We obtain the following results:
\begin{enumerate}
\item We find a strong planet-metallicity correlation for the secure
planets in our Lick sample, with a power law exponent of {\bfa 
$1.7^{+0.3}_{-0.4}$} that matches the planet-metallicity
correlations derived for main-sequence stars \citep{fischer05,udry07}.
\item Giant planet occurrence rate and stellar mass are strongly correlated.
Based on a Gaussian fit the planet occurrence rate is highest for a stellar
mass of {\bfa $1.9^{+0.1}_{-0.5}$\,\msun}, and decreases rapidly for stars with masses
larger than 2.5--3.0\,\msun. The width $\sigma$ of the Gaussian distribution is
{\bfa $0.5^{+0.5}_{-0.2}$\,\msun}. 
The giant planet occurrence rate in the stellar mass interval
from 2.7 to 5.0\,\msun\ is $<1.6$\% with 68.3\% confidence. 
\item We repeat the statistical analysis on a subsample of our full Lick
sample which ensures uniform planet detection capability (in terms of 
number of observations, stellar jitter and stellar mass). 
Since we obtain the same results for the subsample as for the full sample
to within the statistical errors, we conclude that our analysis 
of the full sample is not severely biased. 
\item We do not find a planet-metallicity correlation for the planet
candidates in our sample. We conclude that at least some of the planet
candidates may not be real, despite clear periodicities in their
radial velocities.
\item A strong decrease in planet occurrence rates for stellar masses above roughly 
1.5 to 3\,\msun\ is predicted in the semi-analytic planet formations models by
\citet{ida05} and \citet{kennedy08b}. The decrease in giant planet occurrence 
rates might be explained by a smaller growth rate, longer migration timescale and 
the shorter lifetime of the protostellar disk for the more massive stars compared 
to the less massive ones. 
\item The observed paucity of giant planets for stars more massive than
2.5--3\,\msun\ is most likely not due to stellar evolutionary effects which
might influence the orbital parameters of orbiting planets. More likely, giant
planets with periods up to several years are not being formed in the first place.

\end{enumerate}

%______________________________________________________________

%\section{Conclusions}

\begin{acknowledgements}
The authors would like to express their gratitude to the UCO/Lick staff for
their enduring support. We also want to thank all the CAT observers
who made this work possible, especially David Mitchell, Saskia Hekker, 
Christian Schwab, Simon Albrecht, David Bauer, Stanley Browne,
Kelsey Clubb, Dennis K\"ugler, Julian St\"urmer, Kirsten Vincke and
Dominika Wylezalek. We are indebted to Geoff Marcy and Debra Fischer, who 
both provided invaluable help especially in the early phases of this project.
\end{acknowledgements}

\bibliographystyle{aa}
\bibliography{kgiant_companions}

\begin{thebibliography}{94}
\expandafter\ifx\csname natexlab\endcsname\relax\def\natexlab#1{#1}\fi

\bibitem[{{Baines} {et~al.}(2011){Baines}, {McAlister}, {ten Brummelaar},
  {Turner}, {Sturmann}, {Sturmann}, {Goldfinger}, {Farrington}, \&
  {Ridgway}}]{baines11}
{Baines}, E.~K., {McAlister}, H.~A., {ten Brummelaar}, T.~A., {et~al.} 2011,
  \apj, 743, 130

\bibitem[{{Baraffe} {et~al.}(2010){Baraffe}, {Chabrier}, \&
  {Barman}}]{baraffe10}
{Baraffe}, I., {Chabrier}, G., \& {Barman}, T. 2010, Reports on Progress in
  Physics, 73, 016901

\bibitem[{{Batalha} {et~al.}(2013){Batalha}, {Rowe}, {Bryson}, {Barclay},
  {Burke}, {Caldwell}, {Christiansen}, {Mullally}, {Thompson}, {Brown},
  {Dupree}, {Fabrycky}, {Ford}, {Fortney}, {Gilliland}, {Isaacson}, {Latham},
  {Marcy}, {Quinn}, {Ragozzine}, {Shporer}, {Borucki}, {Ciardi}, {Gautier},
  {Haas}, {Jenkins}, {Koch}, {Lissauer}, {Rapin}, {Basri}, {Boss}, {Buchhave},
  {Carter}, {Charbonneau}, {Christensen-Dalsgaard}, {Clarke}, {Cochran},
  {Demory}, {Desert}, {Devore}, {Doyle}, {Esquerdo}, {Everett}, {Fressin},
  {Geary}, {Girouard}, {Gould}, {Hall}, {Holman}, {Howard}, {Howell},
  {Ibrahim}, {Kinemuchi}, {Kjeldsen}, {Klaus}, {Li}, {Lucas}, {Meibom},
  {Morris}, {Pr{\v s}a}, {Quintana}, {Sanderfer}, {Sasselov}, {Seader},
  {Smith}, {Steffen}, {Still}, {Stumpe}, {Tarter}, {Tenenbaum}, {Torres},
  {Twicken}, {Uddin}, {Van Cleve}, {Walkowicz}, \& {Welsh}}]{batalha13}
{Batalha}, N.~M., {Rowe}, J.~F., {Bryson}, S.~T., {et~al.} 2013, \apjs, 204, 24

\bibitem[{{Bedding} {et~al.}(2010){Bedding}, {Huber}, {Stello}, {Elsworth},
  {Hekker}, {Kallinger}, {Mathur}, {Mosser}, {Preston}, {Ballot}, {Barban},
  {Broomhall}, {Buzasi}, {Chaplin}, {Garc{\'{\i}}a}, {Gruberbauer}, {Hale}, {De
  Ridder}, {Frandsen}, {Borucki}, {Brown}, {Christensen-Dalsgaard},
  {Gilliland}, {Jenkins}, {Kjeldsen}, {Koch}, {Belkacem}, {Bildsten}, {Bruntt},
  {Campante}, {Deheuvels}, {Derekas}, {Dupret}, {Goupil}, {Hatzes}, {Houdek},
  {Ireland}, {Jiang}, {Karoff}, {Kiss}, {Lebreton}, {Miglio}, {Montalb{\'a}n},
  {Noels}, {Roxburgh}, {Sangaralingam}, {Stevens}, {Suran}, {Tarrant}, \&
  {Weiss}}]{bedding10}
{Bedding}, T.~R., {Huber}, D., {Stello}, D., {et~al.} 2010, \apjl, 713, L176

\bibitem[{{Borucki} {et~al.}(2011{\natexlab{a}}){Borucki}, {Koch}, {Basri},
  {Batalha}, {Boss}, {Brown}, {Caldwell}, {Christensen-Dalsgaard}, {Cochran},
  {DeVore}, {Dunham}, {Dupree}, {Gautier}, {Geary}, {Gilliland}, {Gould},
  {Howell}, {Jenkins}, {Kjeldsen}, {Latham}, {Lissauer}, {Marcy}, {Monet},
  {Sasselov}, {Tarter}, {Charbonneau}, {Doyle}, {Ford}, {Fortney}, {Holman},
  {Seager}, {Steffen}, {Welsh}, {Allen}, {Bryson}, {Buchhave},
  {Chandrasekaran}, {Christiansen}, {Ciardi}, {Clarke}, {Dotson}, {Endl},
  {Fischer}, {Fressin}, {Haas}, {Horch}, {Howard}, {Isaacson}, {Kolodziejczak},
  {Li}, {MacQueen}, {Meibom}, {Prsa}, {Quintana}, {Rowe}, {Sherry},
  {Tenenbaum}, {Torres}, {Twicken}, {Van Cleve}, {Walkowicz}, \&
  {Wu}}]{borucki11a}
{Borucki}, W.~J., {Koch}, D.~G., {Basri}, G., {et~al.} 2011{\natexlab{a}},
  \apj, 728, 117

\bibitem[{{Borucki} {et~al.}(2011{\natexlab{b}}){Borucki}, {Koch}, {Basri},
  {Batalha}, {Brown}, {Bryson}, {Caldwell}, {Christensen-Dalsgaard}, {Cochran},
  {DeVore}, {Dunham}, {Gautier}, {Geary}, {Gilliland}, {Gould}, {Howell},
  {Jenkins}, {Latham}, {Lissauer}, {Marcy}, {Rowe}, {Sasselov}, {Boss},
  {Charbonneau}, {Ciardi}, {Doyle}, {Dupree}, {Ford}, {Fortney}, {Holman},
  {Seager}, {Steffen}, {Tarter}, {Welsh}, {Allen}, {Buchhave}, {Christiansen},
  {Clarke}, {Das}, {D{\'e}sert}, {Endl}, {Fabrycky}, {Fressin}, {Haas},
  {Horch}, {Howard}, {Isaacson}, {Kjeldsen}, {Kolodziejczak}, {Kulesa}, {Li},
  {Lucas}, {Machalek}, {McCarthy}, {MacQueen}, {Meibom}, {Miquel}, {Prsa},
  {Quinn}, {Quintana}, {Ragozzine}, {Sherry}, {Shporer}, {Tenenbaum}, {Torres},
  {Twicken}, {Van Cleve}, {Walkowicz}, {Witteborn}, \& {Still}}]{borucki11b}
{Borucki}, W.~J., {Koch}, D.~G., {Basri}, G., {et~al.} 2011{\natexlab{b}},
  \apj, 736, 19

\bibitem[{{Burkert} \& {Ida}(2007)}]{burkert07}
{Burkert}, A. \& {Ida}, S. 2007, \apj, 660, 845

\bibitem[{{Butler} {et~al.}(1996){Butler}, {Marcy}, {Williams}, {McCarthy},
  {Dosanjh}, \& {Vogt}}]{butler96}
{Butler}, R.~P., {Marcy}, G.~W., {Williams}, E., {et~al.} 1996, \pasp, 108, 500

\bibitem[{{Cameron}(2011)}]{cameron11}
{Cameron}, E. 2011, \pasa, 28, 128

\bibitem[{{Carrier} {et~al.}(2010){Carrier}, {De Ridder}, {Baudin}, {Barban},
  {Hatzes}, {Hekker}, {Kallinger}, {Miglio}, {Montalb{\'a}n}, {Morel}, {Weiss},
  {Auvergne}, {Baglin}, {Catala}, {Michel}, \& {Samadi}}]{carrier10}
{Carrier}, F., {De Ridder}, J., {Baudin}, F., {et~al.} 2010, \aap, 509, A73

\bibitem[{{De Ridder} {et~al.}(2009){De Ridder}, {Barban}, {Baudin}, {Carrier},
  {Hatzes}, {Hekker}, {Kallinger}, {Weiss}, {Baglin}, {Auvergne}, {Samadi},
  {Barge}, \& {Deleuil}}]{deridder09}
{De Ridder}, J., {Barban}, C., {Baudin}, F., {et~al.} 2009, \nat, 459, 398

\bibitem[{{D{\"o}llinger} {et~al.}(2009{\natexlab{a}}){D{\"o}llinger},
  {Hatzes}, {Pasquini}, {Guenther}, \& {Hartmann}}]{doellinger09b}
{D{\"o}llinger}, M.~P., {Hatzes}, A.~P., {Pasquini}, L., {Guenther}, E.~W., \&
  {Hartmann}, M. 2009{\natexlab{a}}, \aap, 505, 1311

\bibitem[{{D{\"o}llinger} {et~al.}(2009{\natexlab{b}}){D{\"o}llinger},
  {Hatzes}, {Pasquini}, {Guenther}, {Hartmann}, \& {Girardi}}]{doellinger09a}
{D{\"o}llinger}, M.~P., {Hatzes}, A.~P., {Pasquini}, L., {et~al.}
  2009{\natexlab{b}}, \aap, 499, 935

\bibitem[{{D{\"o}llinger} {et~al.}(2007){D{\"o}llinger}, {Hatzes}, {Pasquini},
  {Guenther}, {Hartmann}, {Girardi}, \& {Esposito}}]{doellinger07}
{D{\"o}llinger}, M.~P., {Hatzes}, A.~P., {Pasquini}, L., {et~al.} 2007, \aap,
  472, 649

\bibitem[{{Fischer} \& {Valenti}(2005)}]{fischer05}
{Fischer}, D.~A. \& {Valenti}, J. 2005, \apj, 622, 1102

\bibitem[{{Fressin} {et~al.}(2013){Fressin}, {Torres}, {Charbonneau}, {Bryson},
  {Christiansen}, {Dressing}, {Jenkins}, {Walkowicz}, \& {Batalha}}]{fressin13}
{Fressin}, F., {Torres}, G., {Charbonneau}, D., {et~al.} 2013, \apj, 766, 81

\bibitem[{{Frink} {et~al.}(2002){Frink}, {Mitchell}, {Quirrenbach}, {Fischer},
  {Marcy}, \& {Butler}}]{frink02}
{Frink}, S., {Mitchell}, D.~S., {Quirrenbach}, A., {et~al.} 2002, \apj, 576,
  478

\bibitem[{{Frink} {et~al.}(2001){Frink}, {Quirrenbach}, {Fischer}, {R{\"o}ser},
  \& {Schilbach}}]{frink01}
{Frink}, S., {Quirrenbach}, A., {Fischer}, D., {R{\"o}ser}, S., \& {Schilbach},
  E. 2001, \pasp, 113, 173

\bibitem[{{Galland} {et~al.}(2006){Galland}, {Lagrange}, {Udry}, {Beuzit},
  {Pepe}, \& {Mayor}}]{galland06}
{Galland}, F., {Lagrange}, A.-M., {Udry}, S., {et~al.} 2006, \aap, 452, 709

\bibitem[{{Galland} {et~al.}(2005{\natexlab{a}}){Galland}, {Lagrange}, {Udry},
  {Chelli}, {Pepe}, {Beuzit}, \& {Mayor}}]{galland05a}
{Galland}, F., {Lagrange}, A.-M., {Udry}, S., {et~al.} 2005{\natexlab{a}},
  \aap, 444, L21

\bibitem[{{Galland} {et~al.}(2005{\natexlab{b}}){Galland}, {Lagrange}, {Udry},
  {Chelli}, {Pepe}, {Queloz}, {Beuzit}, \& {Mayor}}]{galland05b}
{Galland}, F., {Lagrange}, A.-M., {Udry}, S., {et~al.} 2005{\natexlab{b}},
  \aap, 443, 337

\bibitem[{{Gettel} {et~al.}(2012{\natexlab{a}}){Gettel}, {Wolszczan},
  {Niedzielski}, {Nowak}, {Adam{\'o}w}, {Zieli{\'n}ski}, \&
  {Maciejewski}}]{gettel12a}
{Gettel}, S., {Wolszczan}, A., {Niedzielski}, A., {et~al.} 2012{\natexlab{a}},
  \apj, 756, 53

\bibitem[{{Gettel} {et~al.}(2012{\natexlab{b}}){Gettel}, {Wolszczan},
  {Niedzielski}, {Nowak}, {Adam{\'o}w}, {Zieli{\'n}ski}, \&
  {Maciejewski}}]{gettel12b}
{Gettel}, S., {Wolszczan}, A., {Niedzielski}, A., {et~al.} 2012{\natexlab{b}},
  \apj, 745, 28

\bibitem[{{Girardi} {et~al.}(2000){Girardi}, {Bressan}, {Bertelli}, \&
  {Chiosi}}]{girardi00}
{Girardi}, L., {Bressan}, A., {Bertelli}, G., \& {Chiosi}, C. 2000, \aaps, 141,
  371

\bibitem[{{Han} {et~al.}(2010){Han}, {Lee}, {Kim}, {Mkrtichian}, {Hatzes}, \&
  {Valyavin}}]{han10}
{Han}, I., {Lee}, B.~C., {Kim}, K.~M., {et~al.} 2010, \aap, 509, A24

\bibitem[{{Hatzes}(2002)}]{hatzes02}
{Hatzes}, A.~P. 2002, Astronomische Nachrichten, 323, 392

\bibitem[{{Hatzes} \& {Cochran}(1999)}]{hatzes99}
{Hatzes}, A.~P. \& {Cochran}, W.~D. 1999, \mnras, 304, 109

\bibitem[{{Hatzes} {et~al.}(2006){Hatzes}, {Cochran}, {Endl}, {Guenther},
  {Saar}, {Walker}, {Yang}, {Hartmann}, {Esposito}, {Paulson}, \&
  {D{\"o}llinger}}]{hatzes06}
{Hatzes}, A.~P., {Cochran}, W.~D., {Endl}, M., {et~al.} 2006, \aap, 457, 335

\bibitem[{{Hatzes} {et~al.}(2005){Hatzes}, {Guenther}, {Endl}, {Cochran},
  {D{\"o}llinger}, \& {Bedalov}}]{hatzes05}
{Hatzes}, A.~P., {Guenther}, E.~W., {Endl}, M., {et~al.} 2005, \aap, 437, 743

\bibitem[{{Hatzes} {et~al.}(2012){Hatzes}, {Zechmeister}, {Matthews},
  {Kuschnig}, {Walker}, {D{\"o}llinger}, {Guenther}, {Moffat}, {Rucinski},
  {Sasselov}, \& {Weiss}}]{hatzes12}
{Hatzes}, A.~P., {Zechmeister}, M., {Matthews}, J., {et~al.} 2012, \aap, 543,
  A98

\bibitem[{{Hekker} \& {Mel{\'e}ndez}(2007)}]{hekker07}
{Hekker}, S. \& {Mel{\'e}ndez}, J. 2007, \aap, 475, 1003

\bibitem[{{Hekker} {et~al.}(2006){Hekker}, {Reffert}, {Quirrenbach},
  {Mitchell}, {Fischer}, {Marcy}, \& {Butler}}]{hekker06}
{Hekker}, S., {Reffert}, S., {Quirrenbach}, A., {et~al.} 2006, \aap, 454, 943

\bibitem[{{Hekker} {et~al.}(2008){Hekker}, {Snellen}, {Aerts}, {Quirrenbach},
  {Reffert}, \& {Mitchell}}]{hekker08}
{Hekker}, S., {Snellen}, I.~A.~G., {Aerts}, C., {et~al.} 2008, \aap, 480, 215

\bibitem[{{Howard} {et~al.}(2012){Howard}, {Marcy}, {Bryson}, {Jenkins},
  {Rowe}, {Batalha}, {Borucki}, {Koch}, {Dunham}, {Gautier}, {Van Cleve},
  {Cochran}, {Latham}, {Lissauer}, {Torres}, {Brown}, {Gilliland}, {Buchhave},
  {Caldwell}, {Christensen-Dalsgaard}, {Ciardi}, {Fressin}, {Haas}, {Howell},
  {Kjeldsen}, {Seager}, {Rogers}, {Sasselov}, {Steffen}, {Basri},
  {Charbonneau}, {Christiansen}, {Clarke}, {Dupree}, {Fabrycky}, {Fischer},
  {Ford}, {Fortney}, {Tarter}, {Girouard}, {Holman}, {Johnson}, {Klaus},
  {Machalek}, {Moorhead}, {Morehead}, {Ragozzine}, {Tenenbaum}, {Twicken},
  {Quinn}, {Isaacson}, {Shporer}, {Lucas}, {Walkowicz}, {Welsh}, {Boss},
  {Devore}, {Gould}, {Smith}, {Morris}, {Prsa}, {Morton}, {Still}, {Thompson},
  {Mullally}, {Endl}, \& {MacQueen}}]{howard12}
{Howard}, A.~W., {Marcy}, G.~W., {Bryson}, S.~T., {et~al.} 2012, \apjs, 201, 15

\bibitem[{{Ida} \& {Lin}(2005)}]{ida05}
{Ida}, S. \& {Lin}, D.~N.~C. 2005, \apj, 626, 1045

\bibitem[{{Ida} \& {Lin}(2008)}]{ida08}
{Ida}, S. \& {Lin}, D.~N.~C. 2008, \apj, 673, 487

\bibitem[{{Johnson} {et~al.}(2010{\natexlab{a}}){Johnson}, {Aller}, {Howard},
  \& {Crepp}}]{johnson10}
{Johnson}, J.~A., {Aller}, K.~M., {Howard}, A.~W., \& {Crepp}, J.~R.
  2010{\natexlab{a}}, \pasp, 122, 905

\bibitem[{{Johnson} {et~al.}(2010{\natexlab{b}}){Johnson}, {Bowler}, {Howard},
  {Henry}, {Marcy}, {Isaacson}, {Brewer}, {Fischer}, {Morton}, \&
  {Crepp}}]{johnson10b}
{Johnson}, J.~A., {Bowler}, B.~P., {Howard}, A.~W., {et~al.}
  2010{\natexlab{b}}, \apjl, 721, L153

\bibitem[{{Johnson} {et~al.}(2007){Johnson}, {Butler}, {Marcy}, {Fischer},
  {Vogt}, {Wright}, \& {Peek}}]{johnson07}
{Johnson}, J.~A., {Butler}, R.~P., {Marcy}, G.~W., {et~al.} 2007, \apj, 670,
  833

\bibitem[{{Johnson} {et~al.}(2014){Johnson}, {Huber}, {Boyajian}, {Brewer},
  {White}, {von Braun}, {Maestro}, {Stello}, \& {Barclay}}]{johnson14}
{Johnson}, J.~A., {Huber}, D., {Boyajian}, T., {et~al.} 2014, \apj, 794, 15

\bibitem[{{Johnson} {et~al.}(2006){Johnson}, {Marcy}, {Fischer}, {Henry},
  {Wright}, {Isaacson}, \& {McCarthy}}]{johnson06}
{Johnson}, J.~A., {Marcy}, G.~W., {Fischer}, D.~A., {et~al.} 2006, \apj, 652,
  1724

\bibitem[{{Johnson} {et~al.}(2013){Johnson}, {Morton}, \& {Wright}}]{johnson13}
{Johnson}, J.~A., {Morton}, T.~D., \& {Wright}, J.~T. 2013, \apj, 763, 53

\bibitem[{{Jones} {et~al.}(2014{\natexlab{a}}){Jones}, {Jenkins}, {Bluhm},
  {Rojo}, \& {Melo}}]{jones14}
{Jones}, M.~I., {Jenkins}, J.~S., {Bluhm}, P., {Rojo}, P., \& {Melo}, C.~H.~F.
  2014{\natexlab{a}}, ArXiv e-prints

\bibitem[{{Jones} {et~al.}(2011){Jones}, {Jenkins}, {Rojo}, \&
  {Melo}}]{jones11}
{Jones}, M.~I., {Jenkins}, J.~S., {Rojo}, P., \& {Melo}, C.~H.~F. 2011, \aap,
  536, A71

\bibitem[{{Jones} {et~al.}(2013){Jones}, {Jenkins}, {Rojo}, {Melo}, \&
  {Bluhm}}]{jones13}
{Jones}, M.~I., {Jenkins}, J.~S., {Rojo}, P., {Melo}, C.~H.~F., \& {Bluhm}, P.
  2013, \aap, 556, A78

\bibitem[{{Jones} {et~al.}(2014{\natexlab{b}}){Jones}, {Jenkins}, {Rojo},
  {Melo}, \& {Bluhm}}]{jones14b}
{Jones}, M.~I., {Jenkins}, J.~S., {Rojo}, P., {Melo}, C.~H.~F., \& {Bluhm}, P.
  2014{\natexlab{b}}, ArXiv e-prints

\bibitem[{{Kallinger} {et~al.}(2012){Kallinger}, {Hekker}, {Mosser}, {De
  Ridder}, {Bedding}, {Elsworth}, {Gruberbauer}, {Guenther}, {Stello}, {Basu},
  {Garc{\'{\i}}a}, {Chaplin}, {Mullally}, {Still}, \& {Thompson}}]{kallinger12}
{Kallinger}, T., {Hekker}, S., {Mosser}, B., {et~al.} 2012, \aap, 541, A51

\bibitem[{{Kass} \& {Raftery}(1995)}]{kass95}
{Kass}, R.~E. \& {Raftery}, A.~E. 1995, J.~Am.~Stat.~Assoc., 90, 773

\bibitem[{{Kennedy} \& {Kenyon}(2008{\natexlab{a}})}]{kennedy08b}
{Kennedy}, G.~M. \& {Kenyon}, S.~J. 2008{\natexlab{a}}, \apj, 682, 1264

\bibitem[{{Kennedy} \& {Kenyon}(2008{\natexlab{b}})}]{kennedy08a}
{Kennedy}, G.~M. \& {Kenyon}, S.~J. 2008{\natexlab{b}}, \apj, 673, 502

\bibitem[{{Kennedy} \& {Kenyon}(2009)}]{kennedy09}
{Kennedy}, G.~M. \& {Kenyon}, S.~J. 2009, \apj, 695, 1210

\bibitem[{{Kunitomo} {et~al.}(2011){Kunitomo}, {Ikoma}, {Sato}, {Katsuta}, \&
  {Ida}}]{kunitomo11}
{Kunitomo}, M., {Ikoma}, M., {Sato}, B., {Katsuta}, Y., \& {Ida}, S. 2011,
  \apj, 737, 66

\bibitem[{{K\"unstler}(2008)}]{kuenstler08}
{K\"unstler}, A. 2008, Massen-und Altersbestimmung einer Auswahl von G- und
  K-Riesensternen, Diplomarbeit, Landessternwarte, Heidelberg University,
  Germany

\bibitem[{{Lagrange} {et~al.}(2009){Lagrange}, {Desort}, {Galland}, {Udry}, \&
  {Mayor}}]{lagrange09}
{Lagrange}, A.-M., {Desort}, M., {Galland}, F., {Udry}, S., \& {Mayor}, M.
  2009, \aap, 495, 335

\bibitem[{{Lee} {et~al.}(2013){Lee}, {Han}, \& {Park}}]{lee13}
{Lee}, B.-C., {Han}, I., \& {Park}, M.-G. 2013, \aap, 549, A2

\bibitem[{{Lee} {et~al.}(2012{\natexlab{a}}){Lee}, {Han}, {Park}, {Mkrtichian},
  \& {Kim}}]{lee12a}
{Lee}, B.-C., {Han}, I., {Park}, M.-G., {Mkrtichian}, D.~E., \& {Kim}, K.-M.
  2012{\natexlab{a}}, \aap, 546, A5

\bibitem[{{Lee} {et~al.}(2012{\natexlab{b}}){Lee}, {Mkrtichian}, {Han}, {Park},
  \& {Kim}}]{lee12b}
{Lee}, B.-C., {Mkrtichian}, D.~E., {Han}, I., {Park}, M.-G., \& {Kim}, K.-M.
  2012{\natexlab{b}}, \aap, 548, A118

\bibitem[{{Lissauer} {et~al.}(2014){Lissauer}, {Marcy}, {Bryson}, {Rowe},
  {Jontof-Hutter}, {Agol}, {Borucki}, {Carter}, {Ford}, {Gilliland}, {Kolbl},
  {Star}, {Steffen}, \& {Torres}}]{lissauer14}
{Lissauer}, J.~J., {Marcy}, G.~W., {Bryson}, S.~T., {et~al.} 2014, \apj, 784,
  44

\bibitem[{{Liu} {et~al.}(2008){Liu}, {Sato}, {Zhao}, {Noguchi}, {Wang},
  {Kambe}, {Ando}, {Izumiura}, {Chen}, {Okada}, {Toyota}, {Omiya}, {Masuda},
  {Takeda}, {Murata}, {Itoh}, {Yoshida}, {Kokubo}, \& {Ida}}]{liu08}
{Liu}, Y., {Sato}, B., {Zhao}, G., {et~al.} 2008, \apj, 672, 553

\bibitem[{{Lloyd}(2011)}]{lloyd11}
{Lloyd}, J.~P. 2011, \apjl, 739, L49

\bibitem[{{Lloyd}(2013)}]{lloyd13}
{Lloyd}, J.~P. 2013, ArXiv 1306.6627

\bibitem[{{Maldonado} {et~al.}(2013){Maldonado}, {Villaver}, \&
  {Eiroa}}]{maldonado13}
{Maldonado}, J., {Villaver}, E., \& {Eiroa}, C. 2013, \aap, 554, A84

\bibitem[{{Mitchell} {et~al.}(2013){Mitchell}, {Reffert}, {Trifonov},
  {Quirrenbach}, \& {Fischer}}]{mitchell13}
{Mitchell}, D.~S., {Reffert}, S., {Trifonov}, T., {Quirrenbach}, A., \&
  {Fischer}, D.~A. 2013, \aap, 555, A87

\bibitem[{{Mortier} {et~al.}(2013){Mortier}, {Santos}, {Sousa}, {Adibekyan},
  {Delgado Mena}, {Tsantaki}, {Israelian}, \& {Mayor}}]{mortier13}
{Mortier}, A., {Santos}, N.~C., {Sousa}, S.~G., {et~al.} 2013, \aap, 557, A70

\bibitem[{{Niedzielski} {et~al.}(2009{\natexlab{a}}){Niedzielski},
  {Go{\'z}dziewski}, {Wolszczan}, {Konacki}, {Nowak}, \&
  {Zieli{\'n}ski}}]{niedzielski09a}
{Niedzielski}, A., {Go{\'z}dziewski}, K., {Wolszczan}, A., {et~al.}
  2009{\natexlab{a}}, \apj, 693, 276

\bibitem[{{Niedzielski} {et~al.}(2007){Niedzielski}, {Konacki}, {Wolszczan},
  {Nowak}, {Maciejewski}, {Gelino}, {Shao}, {Shetrone}, \&
  {Ramsey}}]{niedzielski07}
{Niedzielski}, A., {Konacki}, M., {Wolszczan}, A., {et~al.} 2007, \apj, 669,
  1354

\bibitem[{{Niedzielski} {et~al.}(2009{\natexlab{b}}){Niedzielski}, {Nowak},
  {Adam{\'o}w}, \& {Wolszczan}}]{niedzielski09b}
{Niedzielski}, A., {Nowak}, G., {Adam{\'o}w}, M., \& {Wolszczan}, A.
  2009{\natexlab{b}}, \apj, 707, 768

\bibitem[{{Pasquini} {et~al.}(2007){Pasquini}, {D{\"o}llinger}, {Weiss},
  {Girardi}, {Chavero}, {Hatzes}, {da Silva}, \& {Setiawan}}]{pasquini07}
{Pasquini}, L., {D{\"o}llinger}, M.~P., {Weiss}, A., {et~al.} 2007, \aap, 473,
  979

\bibitem[{{Quirrenbach} {et~al.}(2011){Quirrenbach}, {Reffert}, \&
  {Bergmann}}]{quirrenbach11}
{Quirrenbach}, A., {Reffert}, S., \& {Bergmann}, C. 2011, in American Institute
  of Physics Conference Series, ed. {S.~Schuh, H.~Drechsel, \& U.~Heber}, Vol.
  1331, 102--109

\bibitem[{{Reffert} {et~al.}(2006){Reffert}, {Quirrenbach}, {Mitchell},
  {Albrecht}, {Hekker}, {Fischer}, {Marcy}, \& {Butler}}]{reffert06}
{Reffert}, S., {Quirrenbach}, A., {Mitchell}, D.~S., {et~al.} 2006, \apj, 652,
  661

\bibitem[{{Rowe} {et~al.}(2014){Rowe}, {Bryson}, {Marcy}, {Lissauer},
  {Jontof-Hutter}, {Mullally}, {Gilliland}, {Issacson}, {Ford}, {Howell},
  {Borucki}, {Haas}, {Huber}, {Steffen}, {Thompson}, {Quintana}, {Barclay},
  {Still}, {Fortney}, {Gautier}, {Hunter}, {Caldwell}, {Ciardi}, {Devore},
  {Cochran}, {Jenkins}, {Agol}, {Carter}, \& {Geary}}]{rowe14}
{Rowe}, J.~F., {Bryson}, S.~T., {Marcy}, G.~W., {et~al.} 2014, \apj, 784, 45

\bibitem[{{Sackmann} {et~al.}(1993){Sackmann}, {Boothroyd}, \&
  {Kraemer}}]{sackmann93}
{Sackmann}, I.-J., {Boothroyd}, A.~I., \& {Kraemer}, K.~E. 1993, \apj, 418, 457

\bibitem[{{Salpeter}(1955)}]{imf}
{Salpeter}, E.~E. 1955, \apj, 121, 161

\bibitem[{{Sato} {et~al.}(2003){Sato}, {Ando}, {Kambe}, {Takeda}, {Izumiura},
  {Masuda}, {Watanabe}, {Noguchi}, {Wada}, {Okada}, {Koyano}, {Maehara},
  {Norimoto}, {Okada}, {Shimizu}, {Uraguchi}, {Yanagisawa}, \&
  {Yoshida}}]{sato03}
{Sato}, B., {Ando}, H., {Kambe}, E., {et~al.} 2003, \apjl, 597, L157

\bibitem[{{Sato} {et~al.}(2008{\natexlab{a}}){Sato}, {Izumiura}, {Toyota},
  {Kambe}, {Ikoma}, {Omiya}, {Masuda}, {Takeda}, {Murata}, {Itoh}, {Ando},
  {Yoshida}, {Kokubo}, \& {Ida}}]{sato08a}
{Sato}, B., {Izumiura}, H., {Toyota}, E., {et~al.} 2008{\natexlab{a}}, \pasj,
  60, 539

\bibitem[{{Sato} {et~al.}(2007){Sato}, {Izumiura}, {Toyota}, {Kambe}, {Takeda},
  {Masuda}, {Omiya}, {Murata}, {Itoh}, {Ando}, {Yoshida}, {Ikoma}, {Kokubo}, \&
  {Ida}}]{sato07}
{Sato}, B., {Izumiura}, H., {Toyota}, E., {et~al.} 2007, \apj, 661, 527

\bibitem[{{Sato} {et~al.}(2012){Sato}, {Omiya}, {Harakawa}, {Izumiura},
  {Kambe}, {Takeda}, {Yoshida}, {Itoh}, {Ando}, {Kokubo}, \& {Ida}}]{sato12}
{Sato}, B., {Omiya}, M., {Harakawa}, H., {et~al.} 2012, \pasj, 64, 135

\bibitem[{{Sato} {et~al.}(2010){Sato}, {Omiya}, {Liu}, {Harakawa}, {Izumiura},
  {Kambe}, {Toyota}, {Murata}, {Lee}, {Masuda}, {Takeda}, {Yoshida}, {Itoh},
  {Ando}, {Kokubo}, {Ida}, {Zhao}, \& {Han}}]{sato10}
{Sato}, B., {Omiya}, M., {Liu}, Y., {et~al.} 2010, \pasj, 62, 1063

\bibitem[{{Sato} {et~al.}(2013){Sato}, {Omiya}, {Wittenmyer}, {Harakawa},
  {Nagasawa}, {Izumiura}, {Kambe}, {Takeda}, {Yoshida}, {Itoh}, {Ando},
  {Kokubo}, \& {Ida}}]{sato13}
{Sato}, B., {Omiya}, M., {Wittenmyer}, R.~A., {et~al.} 2013, \apj, 762, 9

\bibitem[{{Sato} {et~al.}(2008{\natexlab{b}}){Sato}, {Toyota}, {Omiya},
  {Izumiura}, {Kambe}, {Masuda}, {Takeda}, {Itoh}, {Ando}, {Yoshida}, {Kokubo},
  \& {Ida}}]{sato08b}
{Sato}, B., {Toyota}, E., {Omiya}, M., {et~al.} 2008{\natexlab{b}}, \pasj, 60,
  1317

\bibitem[{{Schlaufman} \& {Winn}(2013)}]{schlaufman13}
{Schlaufman}, K.~C. \& {Winn}, J.~N. 2013, ArXiv 1306.0567

\bibitem[{{Schneider} {et~al.}(2011){Schneider}, {Dedieu}, {Le Sidaner},
  {Savalle}, \& {Zolotukhin}}]{schneider11}
{Schneider}, J., {Dedieu}, C., {Le Sidaner}, P., {Savalle}, R., \&
  {Zolotukhin}, I. 2011, \aap, 532, A79

\bibitem[{{Schr{\"o}der} \& {Connon Smith}(2008)}]{schroeder08}
{Schr{\"o}der}, K.-P. \& {Connon Smith}, R. 2008, \mnras, 386, 155

\bibitem[{{Setiawan} {et~al.}(2003){Setiawan}, {Pasquini}, {da Silva}, {von der
  L{\"u}he}, \& {Hatzes}}]{setiawan03}
{Setiawan}, J., {Pasquini}, L., {da Silva}, L., {von der L{\"u}he}, O., \&
  {Hatzes}, A. 2003, \aap, 397, 1151

\bibitem[{{Setiawan} {et~al.}(2005){Setiawan}, {Rodmann}, {da Silva}, {Hatzes},
  {Pasquini}, {von der L{\"u}he}, {de Medeiros}, {D{\"o}llinger}, \&
  {Girardi}}]{setiawan05}
{Setiawan}, J., {Rodmann}, J., {da Silva}, L., {et~al.} 2005, \aap, 437, L31

\bibitem[{{Spiegel} {et~al.}(2011){Spiegel}, {Burrows}, \&
  {Milsom}}]{spiegel11}
{Spiegel}, D.~S., {Burrows}, A., \& {Milsom}, J.~A. 2011, \apj, 727, 57

\bibitem[{{Stello} {et~al.}(2013){Stello}, {Huber}, {Bedding}, {Benomar},
  {Bildsten}, {Elsworth}, {Gilliland}, {Mosser}, {Paxton}, \&
  {White}}]{stello13}
{Stello}, D., {Huber}, D., {Bedding}, T.~R., {et~al.} 2013, \apjl, 765, L41

\bibitem[{{Takeda} {et~al.}(2008){Takeda}, {Sato}, \& {Murata}}]{takeda08}
{Takeda}, Y., {Sato}, B., \& {Murata}, D. 2008, \pasj, 60, 781

\bibitem[{{Trifonov} {et~al.}(2014){Trifonov}, {Reffert}, {Tan}, {Lee}, \&
  {Quirrenbach}}]{trifonov14}
{Trifonov}, T., {Reffert}, S., {Tan}, X., {Lee}, M.~H., \& {Quirrenbach}, A.
  2014, \aap, 568, A64

\bibitem[{{Udry} \& {Santos}(2007)}]{udry07}
{Udry}, S. \& {Santos}, N.~C. 2007, \araa, 45, 397

\bibitem[{{Villaver} \& {Livio}(2009)}]{villaver09}
{Villaver}, E. \& {Livio}, M. 2009, \apjl, 705, L81

\bibitem[{{Vogt}(1987)}]{vogt87}
{Vogt}, S.~S. 1987, \pasp, 99, 1214

\bibitem[{{Wright} {et~al.}(2011){Wright}, {Fakhouri}, {Marcy}, {Han}, {Feng},
  {Johnson}, {Howard}, {Fischer}, {Valenti}, {Anderson}, \&
  {Piskunov}}]{wright11}
{Wright}, J.~T., {Fakhouri}, O., {Marcy}, G.~W., {et~al.} 2011, \pasp, 123, 412

\bibitem[{{Zechmeister} {et~al.}(2008){Zechmeister}, {Reffert}, {Hatzes},
  {Endl}, \& {Quirrenbach}}]{zechmeister08}
{Zechmeister}, M., {Reffert}, S., {Hatzes}, A.~P., {Endl}, M., \&
  {Quirrenbach}, A. 2008, \aap, 491, 531

\end{thebibliography}

\listofobjects

\onecolumn
%\onllongtabL{1}{
\begin{landscape}
\begin{longtable}{l c c r r r r r r c c r r c c c c c l}
\caption{\label{allstars} Stellar mass $M_*$, metallicity $[\mathrm{Fe/H}]$, effective temperature $T_{\mathrm{eff}}$,
luminosity $L$,  
surface gravity $\log g$, radius $R_*$, integrated mass loss $M_{\mathrm{loss}}$ and age $t_*$ for all 373 giant stars of the Lick sample. 
The last column indicates whether the parameters refer to the star being on the red giant branch (RGB) 
or horizontal branch (HB), respectively, while the second but last column gives the corresponding probability $P$.
The $\dagger$ symbol marks the stars for which our interpolation did not yield any results.}\\
\hline\hline\noalign{\smallskip}
Identifier & \multicolumn{1}{c}{$M_*$} & \multicolumn{1}{c}{$\sigma_{M_*}$} & \multicolumn{1}{c}{$[\mathrm{Fe/H}]$} & \multicolumn{1}{c}{$\sigma_{[\mathrm{Fe/H}]}$} & \multicolumn{1}{c}{$T_{\mathrm{eff}}$} & \multicolumn{1}{c}{$\sigma_{T_{\mathrm{eff}}}$} & \multicolumn{1}{c}{$L$} & \multicolumn{1}{c}{$\sigma_L$} & \multicolumn{1}{c}{$\log g$} & \multicolumn{1}{c}{$\sigma_{\log g}$} & \multicolumn{1}{c}{$R_*$} & \multicolumn{1}{c}{$\sigma_{R_*}$} & $M_{\mathrm{loss}}$ & $\sigma_{M_{\mathrm{loss}}}$ & $t_*$ & $\sigma_{t_*}$ & $P$ & RGB/ \\
& \multicolumn{1}{c}{$[\mathrm{M_{\sun}}]$} & \multicolumn{1}{c}{$[\mathrm{M_{\sun}}]$} & \multicolumn{1}{c}{$[\mathrm{dex}]$} & \multicolumn{1}{c}{$[\mathrm{dex}]$} & \multicolumn{1}{c}{$[\mathrm{K}]$} & \multicolumn{1}{c}{$[\mathrm{K}]$} & \multicolumn{1}{c}{$[\mathrm{L_{\sun}}]$} & \multicolumn{1}{c}{$[\mathrm{L_{\sun}}]$} & \multicolumn{1}{c}{$[\mathrm{cm/s^2}]$} & \multicolumn{1}{c}{$[\mathrm{cm/s^2}]$} & \multicolumn{1}{c}{$[\mathrm{R_{\sun}}]$} & \multicolumn{1}{c}{$[\mathrm{R_{\sun}}]$} & $[\mathrm{M_{\sun}}]$ & $[\mathrm{M_{\sun}}]$ & [Gyr] & [Gyr] & & HB \\[0.8ex] 
\hline\noalign{\smallskip}
\endfirsthead
\caption{continued.}\\
\hline\hline\noalign{\smallskip}
Identifier & \multicolumn{1}{c}{$M_*$} & \multicolumn{1}{c}{$\sigma_{M_*}$} & \multicolumn{1}{c}{$[\mathrm{Fe/H}]$} & \multicolumn{1}{c}{$\sigma_{[\mathrm{Fe/H}]}$} & \multicolumn{1}{c}{$T_{\mathrm{eff}}$} & \multicolumn{1}{c}{$\sigma_{T_{\mathrm{eff}}}$} & \multicolumn{1}{c}{$L$} & \multicolumn{1}{c}{$\sigma_L$} & \multicolumn{1}{c}{$\log g$} & \multicolumn{1}{c}{$\sigma_{\log g}$} & \multicolumn{1}{c}{$R_*$} & \multicolumn{1}{c}{$\sigma_{R_*}$} & $M_{\mathrm{loss}}$ & $\sigma_{M_{\mathrm{loss}}}$ & $t_*$ & $\sigma_{t_*}$ & $P$ & RGB/\\
& \multicolumn{1}{c}{$[\mathrm{M_{\sun}}]$} & \multicolumn{1}{c}{$[\mathrm{M_{\sun}}]$} & \multicolumn{1}{c}{$[\mathrm{dex}]$} & \multicolumn{1}{c}{$[\mathrm{dex}]$} & \multicolumn{1}{c}{$[\mathrm{K}]$} & \multicolumn{1}{c}{$[\mathrm{K}]$} & \multicolumn{1}{c}{$[\mathrm{L_{\sun}}]$} & \multicolumn{1}{c}{$[\mathrm{L_{\sun}}]$} & \multicolumn{1}{c}{$[\mathrm{cm/s^2}]$} & \multicolumn{1}{c}{$[\mathrm{cm/s^2}]$} & \multicolumn{1}{c}{$[\mathrm{R_{\sun}}]$} & \multicolumn{1}{c}{$[\mathrm{R_{\sun}}]$} & $[\mathrm{M_{\sun}}]$ & $[\mathrm{M_{\sun}}]$ & [Gyr] & [Gyr] & & HB \\[0.8ex]
\hline\noalign{\smallskip}
\endhead
\hline\noalign{\smallskip}
\endfoot
\object{   HIP 379}           &  1.80 &  0.25 & $-0.07$ &  0.10 & 4732 &   32 &    53.1 &     2.9 &   2.63 &  0.08  &  10.86 &  0.33 & 0.01 & 0.00 &  2.37 & 1.05 & 1.00 & RGB\\[0.8ex]
\object{   HIP 476}           &  2.64 &  0.10 & $-0.22$ &  0.06 & 5065 &   21 &    80.2 &     5.0 &   2.74 &  0.03  &  11.65 &  0.37 & 0.01 & 0.00 &  0.49 & 0.04 & 1.00 & RGB\\[0.8ex]
\object{  HIP 1354}           &  3.28 &  0.18 & $-0.26$ &  0.10 & 5068 &   35 &   186.3 &    27.0 &   2.48 &  0.06  &  17.74 &  1.31 & 0.01 & 0.00 &  0.28 & 0.04 & 0.19 & RGB\\
                              &  2.93 &  0.19 & $-0.26$ &  0.10 & 5030 &   15 &   191.8 &    24.9 &   2.40 &  0.05  &  18.27 &  1.19 & 0.01 & 0.00 &  0.43 & 0.10 & 0.81 & HB\\[0.8ex]
\object{  HIP 1562}           &  3.46 &  0.17 & $ 0.03$ &  0.10 & 4487 &   18 &   354.8 &    11.0 &   1.99 &  0.03  &  31.22 &  0.55 & 0.01 & 0.00 &  0.27 & 0.03 & 0.08 & RGB\\
                              &  3.35 &  0.33 & $ 0.03$ &  0.10 & 4494 &   30 &   352.6 &    11.7 &   1.99 &  0.06  &  31.04 &  0.67 & 0.02 & 0.01 &  0.36 & 0.17 & 0.92 & HB\\[0.8ex]
\object{  HIP 2006}           &  2.93 &  0.10 & $-0.13$ &  0.10 & 5210 &   39 &   100.0 &    11.9 &   2.75 &  0.04  &  12.30 &  0.75 & 0.01 & 0.00 &  0.38 & 0.03 & 1.00 & RGB\\[0.8ex]
\object{  HIP 2497}           &  1.68 &  0.20 & $-0.07$ &  0.10 & 4509 &   21 &    95.9 &     9.0 &   2.26 &  0.06  &  16.08 &  0.77 & 0.01 & 0.00 &  3.04 & 1.38 & 0.96 & RGB\\
\end{longtable}
\end{landscape}
%}

\twocolumn

\end{document}